\documentclass[preprint, 3p, twocolumn, times]{elsarticle} 

\journal{NIM A}

\usepackage{heppennames2}
\usepackage[noabbrev,capitalise]{cleveref}
\usepackage{subcaption}
\usepackage[alsoload=hep]{siunitx}
\usepackage{booktabs}

\usepackage[eulergreek]{sansmath}

\usepackage{todonotes}
\usepackage{multirow}
\usepackage{tikz}

 \usepackage{pgfplots} 
 \usepgfplotslibrary{groupplots}

\begin{document}
\begin{frontmatter}
  
\title{Performance evaluation of thin active-edge planar sensors for the CLIC vertex detector\tnoteref{mytitlenote}}
\tnotetext[mytitlenote]{This work was carried out in the framework of the CLICdp collaboration.}

\date{\today}

\author[cern]{N.~Alipour Tehrani}
\author[unige]{M.~Benoit}
\author[cern]{D.~Dannheim}
\author[cern]{A.~Fiergolski}
\author[cern]{D.~Hynds\fnref{fn2}}
\author[cern]{W.~Klempt}
\author[mpi]{A.~Macchiolo\fnref{fn3}}
\author[cern]{M.~Munker}
\author[cern]{A.~N\"urnberg \corref{cor1}\fnref{fn1}}
\ead{andreas.nurnberg@cern.ch}

\cortext[cor1]{Corresponding author}
\address[cern]{CERN, Switzerland}
\address[unige]{University of Geneva, Switzerland}
\address[mpi]{Max-Planck-Institut for Physics, Munich, Germany}


\fntext[fn1]{Now at: KIT, Karlsruhe, Germany.}
\fntext[fn2]{Now at: NIKHEF, Amsterdam, The Netherlands.}
\fntext[fn3]{Now at: University of Zurich, Switzerland.}

\begin{abstract}Thin planar silicon sensors with a pitch of \SI{55}{\micron}, active edge and various guard ring layouts are investigated, using two-dimensional finite-element T-CAD simulations.
The simulation results have been compared to experimental data, and an overall good agreement is observed. It is demonstrated that \SI{50}{\micron} thin planar silicon sensors with active edge with floating guard ring or without guard ring can be operated fully efficiently up to the physical edge of the sensor.
The simulation findings are used to identify suitable sensor designs for application in the high-precision vertex detector of the future CLIC linear \Pep{}\Pem{} collider.
\end{abstract}
\end{frontmatter}

\section{Introduction}
The physics aims at the future \Pep{}\Pem{} Compact Linear Collider CLIC pose challenging requirements on the vertex detector~\cite{cdrvol2}.
Among others, hybrid planar silicon pixel detectors are under study.
The strict limit on the material budget of only \SI{0.2}{\percent{}$X_0$} per detector layer limits the active silicon thickness to only \SI{50}{\micron}.
Active-edge sensors can help to reduce the material content of the detector, by avoiding overlaps between neighbouring sensor tiles and still providing a good coverage of the solid angle.

Thin, active-edge sensors with square pixels with \SI{55}{\micron} pitch have been designed, fabricated and characterised in laboratory and test-beam studies, using Timepix3~\cite{timepix3_paper} readout chips as test vehicle.
Different edge termination and guard ring schemes have been implemented.
Besides the evaluation of the tracking performance, the main focus of the study has been on breakdown behaviour and charge collection efficiency in the edge region.

The detection efficiency of planar silicon pixel sensors with implanted and activated edges is strongly dependent on the electric field distribution in the edge region and hence on the interplay between pixel geometry, guard ring grounding scheme and edge implant. Optimisation of the edge design, in particular of the guard ring structure and the distance of the pixels to the trench around the sensor perimeter is therefore of utmost importance. Therefore, the study is complemented by finite-element T-CAD device simulations. 

\section{Active-edge sensors and simulation setup}

In order to achieve full area coverage in the experiment, non-active regions have to be minimised. Active-edge processing is a technology aiming at eliminating the inactive silicon at the sensor edge, offering the possibility of seamlessly tiling individual sensors without the need for overlaps~\cite{advacam}. Using a deep reactive ion-etching process, a trench at the sensor edge is formed. Ion implantation of the walls results in an extension of the backside electrode to the edge of the sensor. Consequently, the depletion zone is extended and the sensor is able to detect particles passing close to its physical edge. After the sensors are separated, the trench becomes the sensor edge and defines the physical size of the sensor. Using this technology, n-in-p sensors matching the Timepix3 layout with a matrix size of \num{256x256} pixels of \SI{55x55}{\micron} size have been produced by Advacam~\cite{advacam2}.
Due to the extension of the backside contact to the edge, the potential gradient between the outermost pixels (close to ground potential) and the sensor edge (bias voltage) makes this type of sensor prone to early breakdown. To smoothen the potential gradient, guard rings are placed between the outermost pixel cell and the sensor edge, surrounding the full pixel matrix. These guard rings can either be connected to ground potential through an additional row of grounded pixel pads on the Timepix3 chip or can be kept floating.

\subsection{Device layout}

\begin{figure}[tbp]
  \centering
  \begin{subfigure}[b]{0.49\linewidth}
    \centering
    \begin{tikzpicture}
      \node[anchor=south west,inner sep=0] (image) at
      (0,0){\includegraphics[width=.9\linewidth,clip,trim=14cm 0 0 0 ]{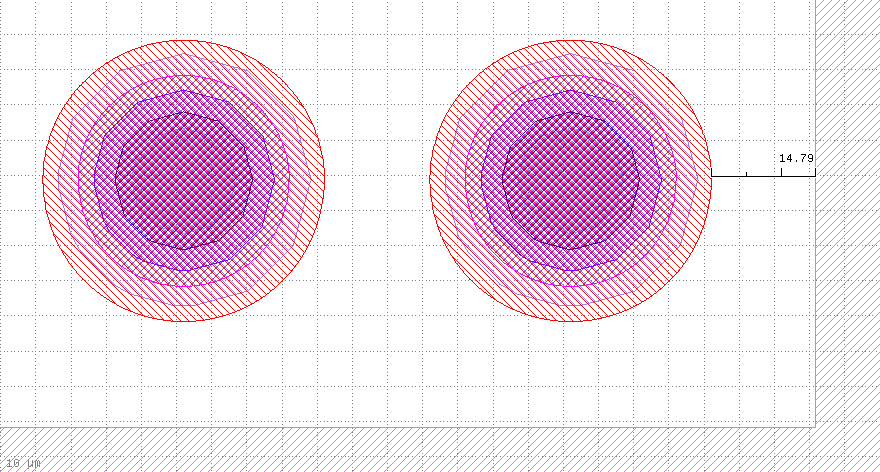}};
      \begin{scope}[x={(image.south east)},y={(image.north west)}]
        \draw[ultra thick,|<->|](0.57,0.6)-- +(0.3,0) node[pos=.5,anchor=south] {\SI{20}{\micron}};
        \draw[<-] (0.4,0.91)-- +(0.2,0.1) node[pos=1, anchor=west]{\scriptsize Metal};
        \draw[<-] (0.4,0.87)-- +(0.2,0.05) node[pos=1, anchor=west]{\scriptsize Etch oxide};
        \draw[<-] (0.4,0.83)-- +(0.2,0.0) node[pos=1, anchor=west]{\scriptsize Pixel implant};
        \draw[<-] (0.4,0.83)-- +(0.2,0.0) node[pos=1, anchor=west]{\scriptsize Pixel implant};
        \draw[<-] (0.4,0.435)-- +(0.2,-0.3) node[pos=1, anchor=west]{\scriptsize UBM};
        \draw[<-] (0.4,0.48)-- +(0.2,-0.2) node[pos=1, anchor=west]{\scriptsize \shortstack{Passiv.\\opening}};
        \draw[<-] (0.4,0.52)-- +(0.2,-0.1) node[pos=1, anchor=west]{\scriptsize Via};
        \draw[<-] (0.5,0.1)-- +(-0.1,0.1) node[pos=1, anchor=east]{\scriptsize \shortstack{Trench/\\physical edge}};
      \end{scope}
    \end{tikzpicture}
    \caption{}
    \label{fig:layout_a}
  \end{subfigure}
  \begin{subfigure}[b]{0.49\linewidth}
    \centering
    \begin{tikzpicture}
      \node[anchor=south west,inner sep=0] (image) at
      (0,0){\includegraphics[width=.9\linewidth,clip,trim=14cm  0 0 0]{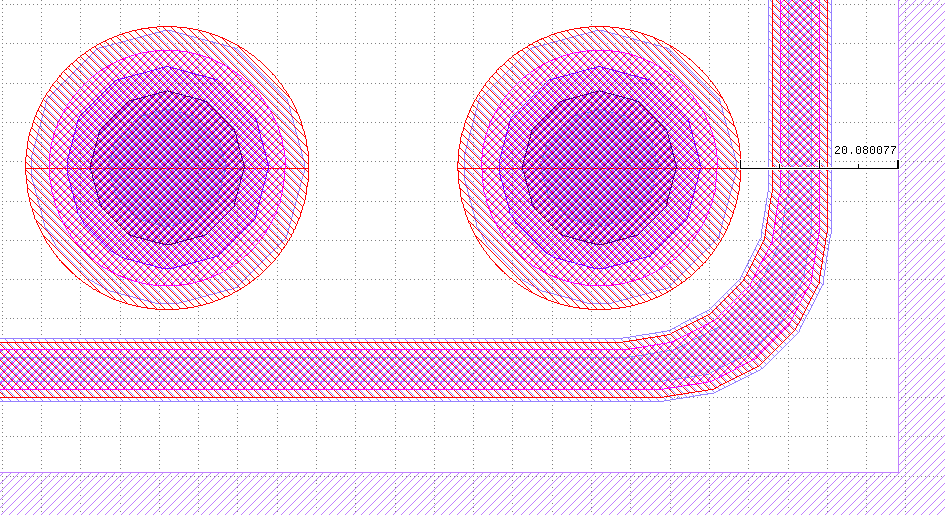}};
      \begin{scope}[x={(image.south east)},y={(image.north west)}]
       \draw[ultra thick,|<->|](0.57,0.675)-- +(0.35,0) node[pos=0.5,anchor=south] {\SI{23}{\micron}};
      \end{scope}
    \end{tikzpicture}
    \caption{}
    \label{fig:layout_b}
  \end{subfigure}

  \caption{Pixel and edge layout of two different investigated sensor designs: (\subref{fig:layout_a}) \SI{20}{\micron} edge distance without guard ring and (\subref{fig:layout_b}) \SI{23}{\micron} edge distance floating guard ring.}
   \label{fig:edge_layouts}
\end{figure}
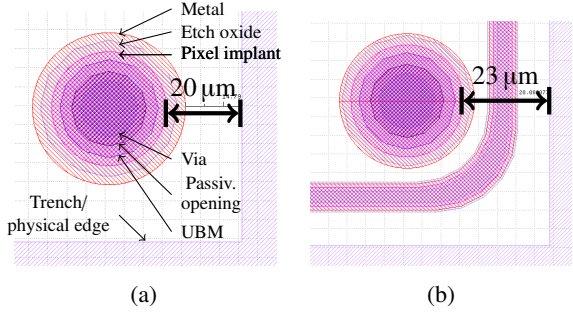

Active-edge sensors with six different guard ring layouts and three different thicknesses between \SIrange[range-phrase={ and }]{50}{150}{\micron} have been investigated. \cref{fig:edge_layouts} depicts two of the different layouts graphically, and \cref{tab:edge_layouts} summarises important parameters of all tested devices, especially the edge distance, which is defined as the distance between the last pixel implant and the physical edge of the sensor.

\begin{table}[tbp]
  \centering
  \caption{Summary of the investigated sensor layouts. Edge distance refers to the distance between the active edge and the last pixel implant.}\label{tab:edge_layouts}
  \begin{tabular}{ c c c c}
    \toprule
    Thickness & Bias voltage & Edge dist. & GR layout\\ \midrule
    \SI{50}{\micron} & \SI{-15}{\volt} & \SI{20}{\micron} & None\\
    \SI{50}{\micron} & \SI{-15}{\volt} & \SI{50}{\micron} & None\\
    \SI{50}{\micron} & \SI{-15}{\volt} & \SI{23}{\micron} & Floating\\
    \SI{50}{\micron} & \SI{-15}{\volt} & \SI{50}{\micron} & Floating\\
    \SI{50}{\micron} & \SI{-15}{\volt} & \SI{28}{\micron} & Grounded\\
    \SI{50}{\micron} & \SI{-15}{\volt} & \SI{55}{\micron} & Grounded\\
    \SI{100}{\micron} & \SI{-20}{\volt} & \SI{55}{\micron} & Grounded\\
    \SI{150}{\micron} & \SI{-30}{\volt} & \SI{55}{\micron} & Grounded\\
    \bottomrule
\end{tabular}
\end{table}

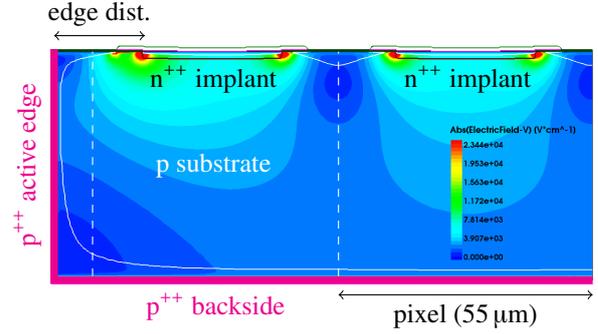
\begin{figure}[t]
  \begin{tikzpicture}
    \node[anchor=south west,inner sep=0] (image) at (0,0)
    {\includegraphics[width=.925\linewidth]{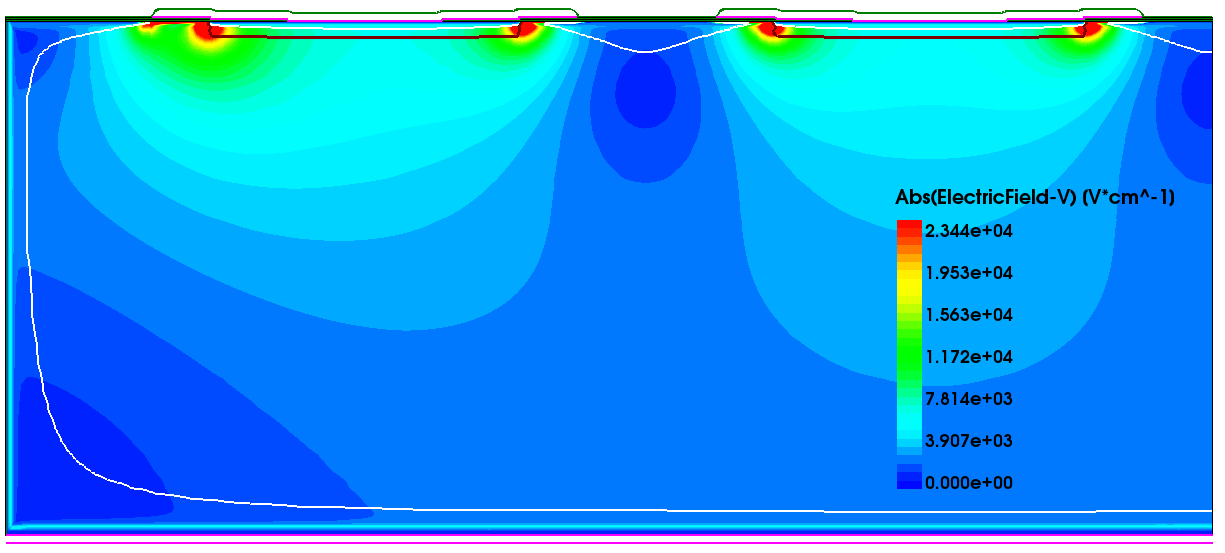}};
    \begin{scope}[x={(image.south east)},y={(image.north west)}]
       \draw[white,dashed](0.5275,0.05) -- ++(0,0.91);
       \draw[white,dashed](0.08,0.05) -- ++(0,0.91);
       \draw[<->](0.5275,-0.025)--(0.99,-0.025) node[below, pos=0.5]{pixel (\SI{55}{\micron})};
       \node[white] at (0.3,0.5){p substrate};
       \node at (0.3,0.85){$\text{n}^{++}$ implant};
       \node at (0.7625,0.85){$\text{n}^{++}$ implant};
       \draw[<->](0.01,1.025) -- (0.175,1.025) node[above, pos=0.5]{edge dist.};
       \draw[line width=1mm, magenta](0.99,0.035)--++(-0.98,0) node[below, pos=.7]{$\text{p}^{++}$ backside} -- ++(0,0.935) node[rotate=90, above, pos=0.5] {$\text{p}^{++}$ active edge};
    \end{scope}
  \end{tikzpicture}
  \caption{Electric field distribution in a \SI{50}{\micron} thick planar sensor at \SI{-15}{\volt} bias voltage in the 2d cross section.
  The solid white line indicates the border of the depletion volume, the dashed white lines indicate the \SI{55}{\micron} pixel grid.}\label{fig:active_edge_sensor}
\end{figure}

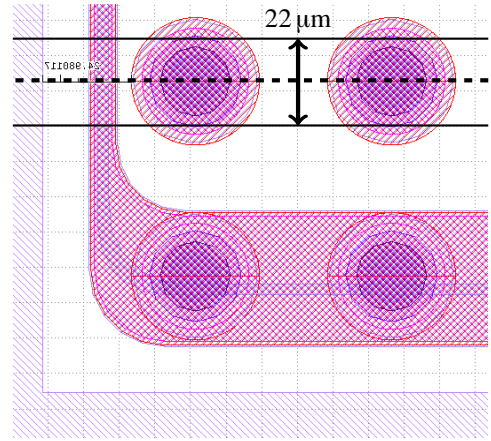
\begin{figure}[t]
  \centering
    \begin{tikzpicture}
      \node[anchor=south west,inner sep=0,xscale=-1]  (image) at
      (0,0){\includegraphics[width=.8\linewidth]{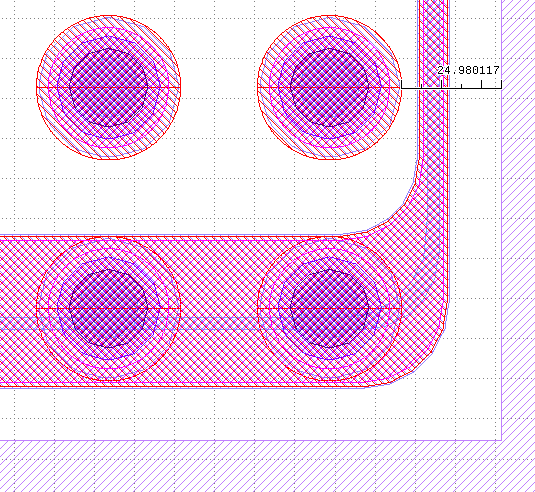}};
      \begin{scope}[x={(image.south east)},y={(image.north west)}]
        \draw[dashed,ultra thick](0,0.825)--+(1,0);
        
        \draw[thick] (0,0.921)--++(1,0);
        \draw [thick](0,0.721)--++(1,0);
        \draw[ultra thick, <->](0.4,0.721)--++(0,0.2) node[pos=1.2]{\SI{22}{\micron}};
      
      \end{scope}
    \end{tikzpicture}
    \caption{Illustration of the simulated cross section through the sensor  layout (dashed line) and geometrical restriction of the reconstructed tracks used for comparing the experimental results with the 2d-simulation (solid lines).}\label{fig:sim_cross_section}
\end{figure}

\cref{fig:sim_cross_section,fig:active_edge_sensor} illustrate the simulated two-dimensional cross section through the sensor layout. Two pixels in neighbouring columns, the optional guard ring and the cut edge are simulated.

\subsection{Experimental setup}

Experimental studies on active-edge sensors have been performed using Timepix3 readout ASICs. The sensors have been operated as Device-Under-Test (DUT) in a Timepix3 reference telescope~\cite{ThesisNilou,lhcb_telescope} in a \SI{120}{\giga\electronvolt} pion beam at the SPS H6 beam line at CERN.\@ The resolution of $\sim\SI{2}{\micron}$ of the reconstructed track position on the DUT allows for a position dependent analysis of the charge collection properties close to the sensor edge. For better comparability of the two-dimensional simulation with the experimental results, only tracks reconstructed within \SI{\pm11}{\micron} around the centre of the pixel row are considered, as illustrated in \cref{fig:sim_cross_section} by the solid lines. 

Current-Voltage (IV) characteristics have been recorded at room temperature using a source-meter. During these measurements, the ASICs were not powered.

\subsection{T-CAD simulation setup}
The simulations are performed using the finite-element T-CAD software Synopsys Sentaurus Device~\cite{synopsys} in version I-2013.12. Both static and transient simulations are performed. At the device boundary, the components of the electric field and carrier currents perpendicular to the surface are set to zero. For contacts connected to external circuitry (backside, pixels and grounded guard rings), the potential is fixed to the external voltage, while for floating contacts (floating guard rings) a zero-current condition is applied. After defining the geometry, the bias voltage on the backside contact is ramped to the nominal value in a quasi-static simulation. From that, the IV characteristics and the electric field distribution in the sensor can be extracted.

In a second step, a particle hit is simulated by introducing the ionisation charges. 
In the HeavyIon model, a given amount of e-h pairs can be distributed in time and space, according to the linear energy transfer of the particle. \SI{500}{\ps} after the start of the time dependent simulation, charge carriers are created along the particle path with a Gaussian profile with width of \SI{0.5}{\micron} and a linear energy transfer of \SI{0.01}{\femto\coulomb\per\micron}, corresponding to about \SI{62}{e-h pairs\per\micron}. The drift of these charges in the sensor volume is tracked in a time-dependent simulation. Simultaneously, the induced current on all electrodes is recorded. 
By integrating the current pulses over the simulation time of \SI{15}{\ns}, the total induced charge is extracted.

\section{Simulation results and comparison with experimental data}
\subsection{Electric field}
\begin{figure}[tbp]
  \begin{subfigure}[T]{.31\linewidth}
    \begin{tikzpicture}
      \node[anchor=south west,inner sep=0] (image) at (0,0)
      {\includegraphics[width=\textwidth,clip=true,trim=3.5cm 4cm 20cm 1cm]{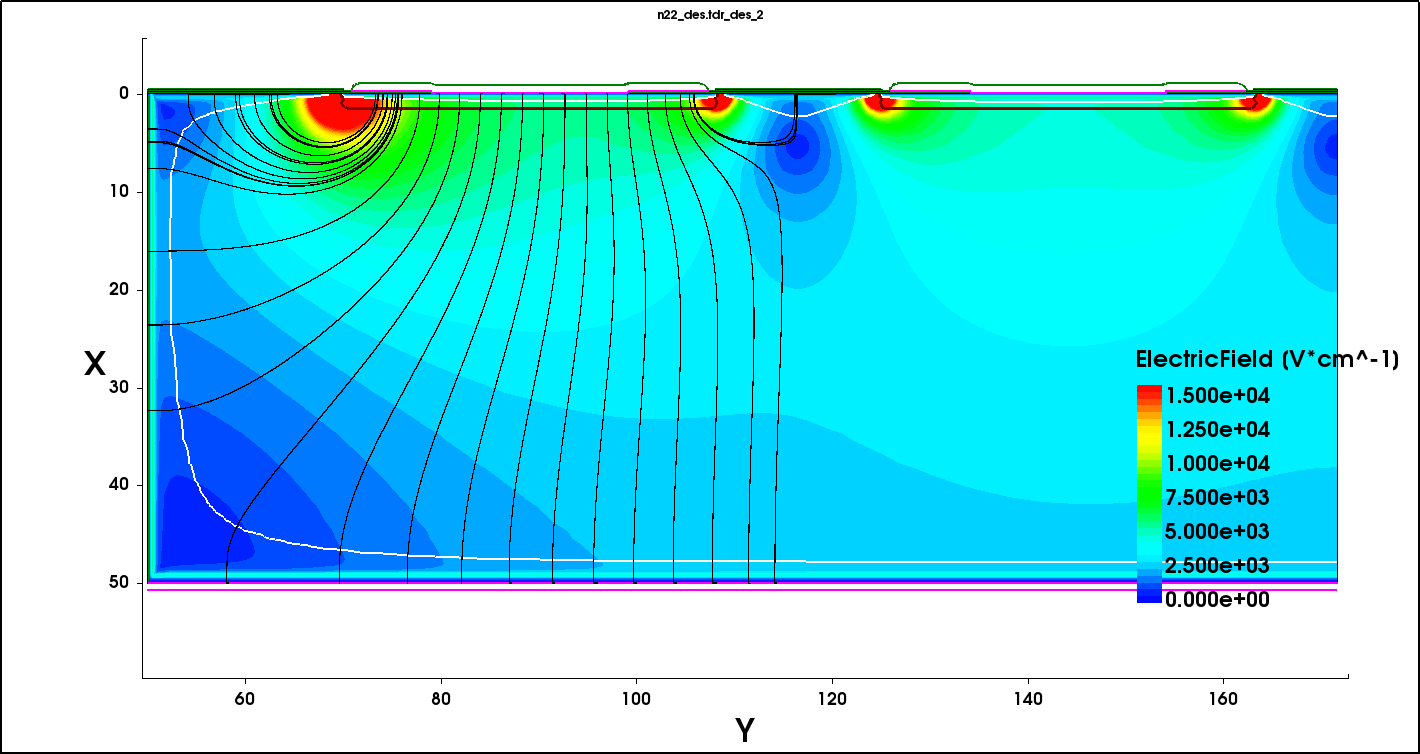}};
      \begin{scope}[x={(image.south east)},y={(image.north west)}]
      \end{scope}
    \end{tikzpicture}
    \caption{No GR}\label{fig:field_nogr}
  \end{subfigure}
  \begin{subfigure}[T]{.31\linewidth}
    \begin{tikzpicture}
      \node[anchor=south west,inner sep=0] (image) at (0,0)
      {\includegraphics[width=\textwidth,clip=true,trim=3.5cm 4cm 20cm 1cm]{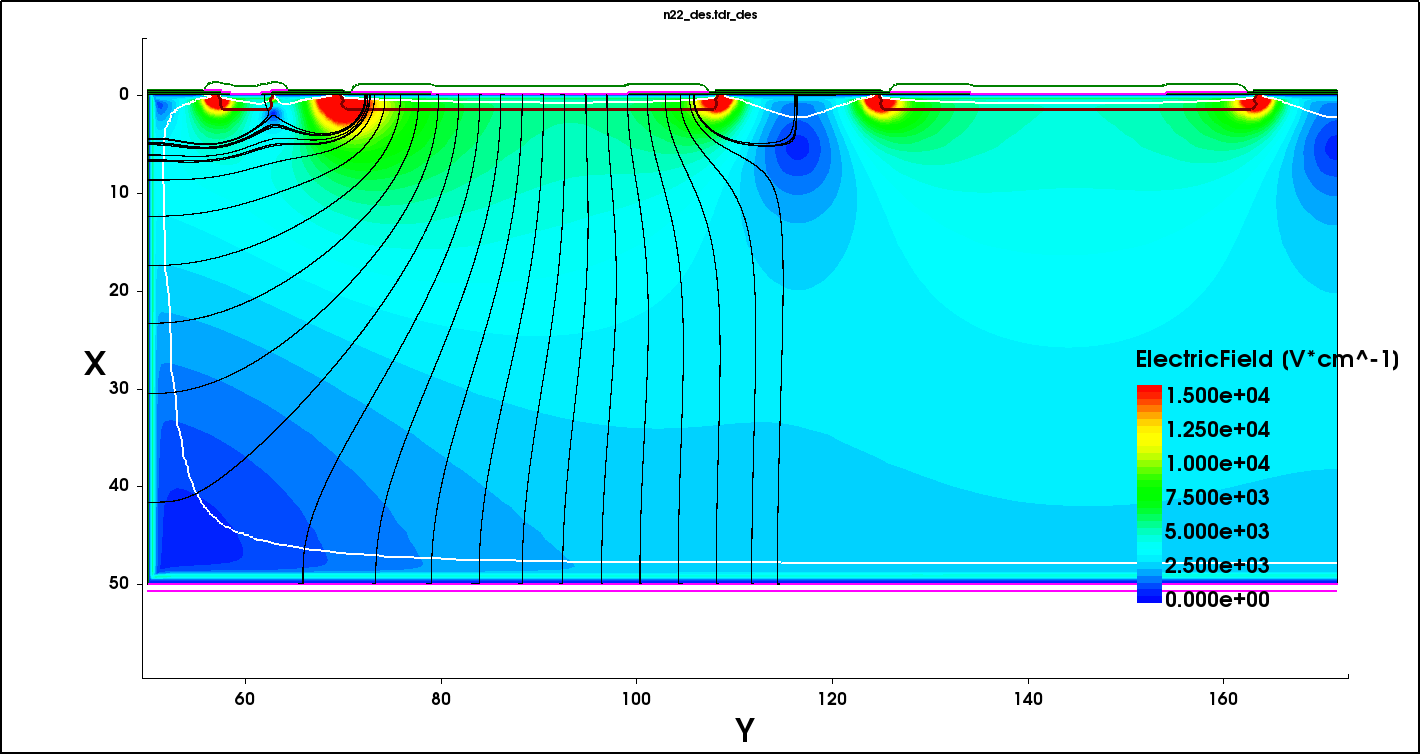}};
      \begin{scope}[x={(image.south east)},y={(image.north west)}]
      \end{scope}
    \end{tikzpicture}
    \caption{Floating GR}\label{fig:field_floatgr}
  \end{subfigure}
  \begin{subfigure}[T]{.31\linewidth}
    \begin{tikzpicture}
      \node[anchor=south west,inner sep=0] (image) at (0,0)
      {\includegraphics[width=\textwidth,clip=true,trim=3.5cm 4cm 20cm 1cm]{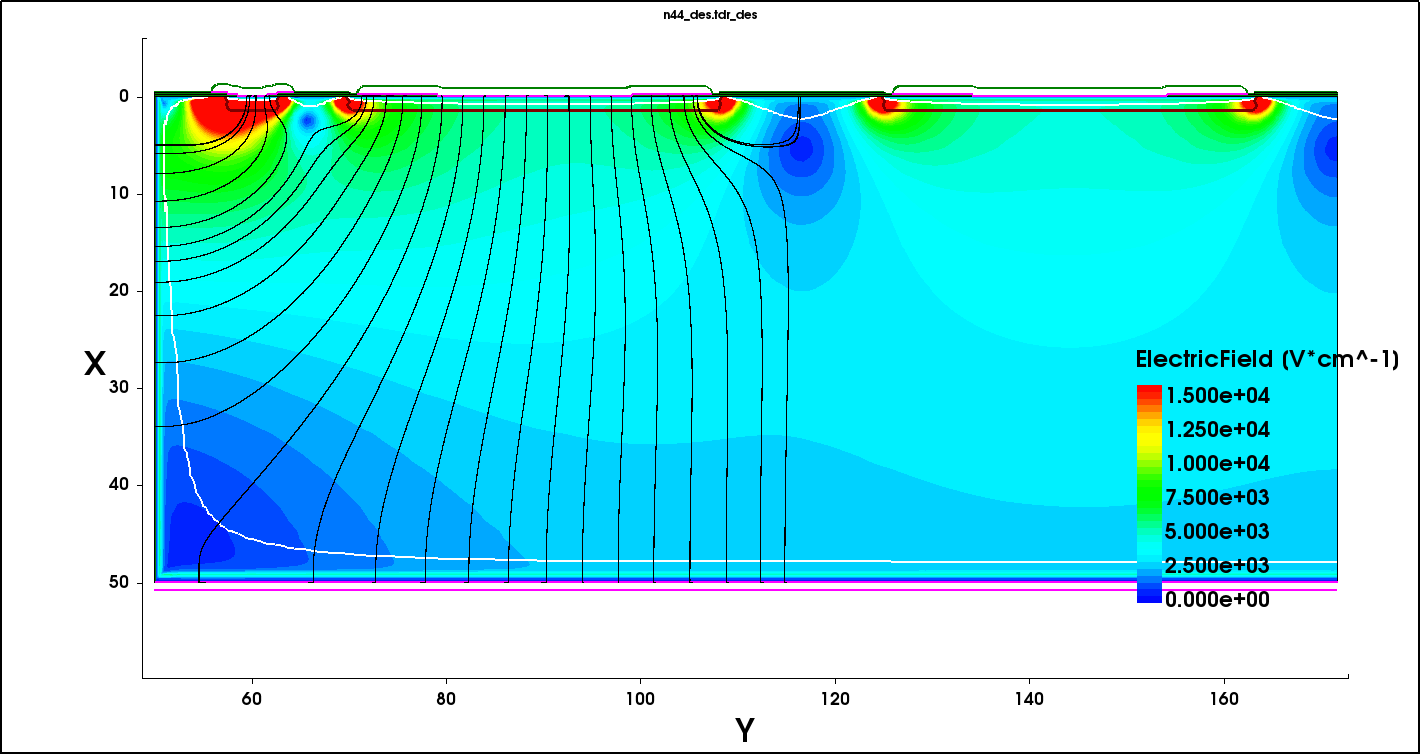}};
      \begin{scope}[x={(image.south east)},y={(image.north west)}]
      \end{scope}
    \end{tikzpicture}
    \caption{Grounded GR}\label{fig:field_gndgr}
  \end{subfigure}
  \caption{Electric field distribution in \SI{50}{\micron} thick sensors close to the edge for different guard ring layouts at \SI{-15}{\volt} bias.
  The solid white line indicates the border of the depletion volume.}\label{fig:efield_50um}
\end{figure}

The electric field configuration close to the sensor edge depends on the guard ring layout, as illustrated in \cref{fig:efield_50um}. Without guard ring and with floating guard ring, the electric field extends directly from the sensor edge to the last pixel row. In devices with floating guard ring, the high-field region between the last pixel and the active edge is reduced. In devices with grounded guard ring, some field lines extend from the sensor edge to the guard ring. This indicates that part of the ionisation charge is collected by the guard ring and the charge collection efficiency of the last pixel will be reduced.

\subsection{IV characteristics and breakdown}
\begin{figure*}[tbp]
  \begin{minipage}{.31\linewidth}
  \includegraphics[width=\linewidth]{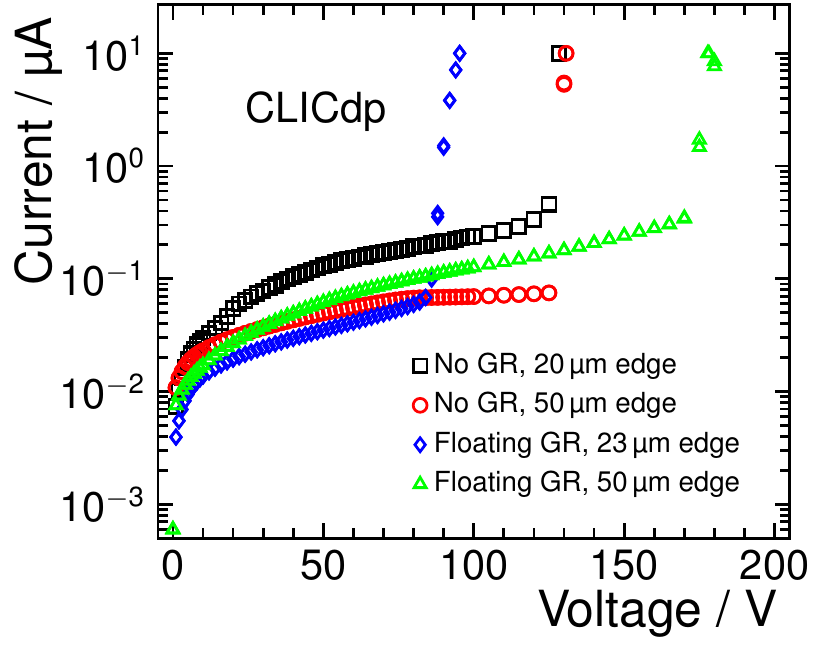}
  \caption{Measured IV characteristics of the  \SI{50}{\micron} thock sensors with different investigated edge termination geometry.}\label{fig:iv_measurement} 
\end{minipage}
\hfill
\begin{minipage}{.31\linewidth}
  \includegraphics[width=\linewidth]{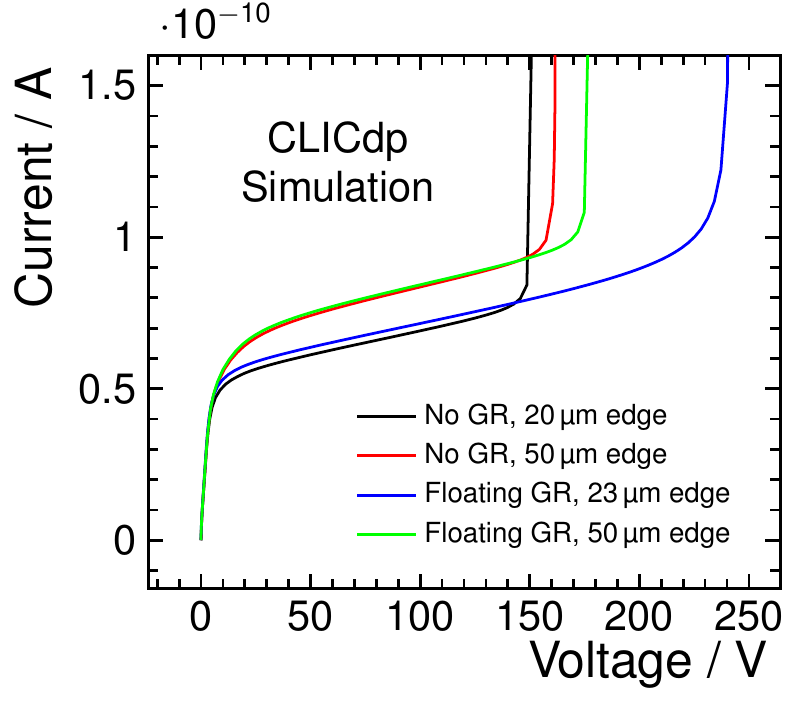}
  \caption{Simulated IV characteristics of the \SI{50}{\micron} thick sensors with different edge layouts.}\label{fig:iv_50um}
\end{minipage}
\hfill
\begin{minipage}{.31\linewidth}
  \includegraphics[width=\linewidth]{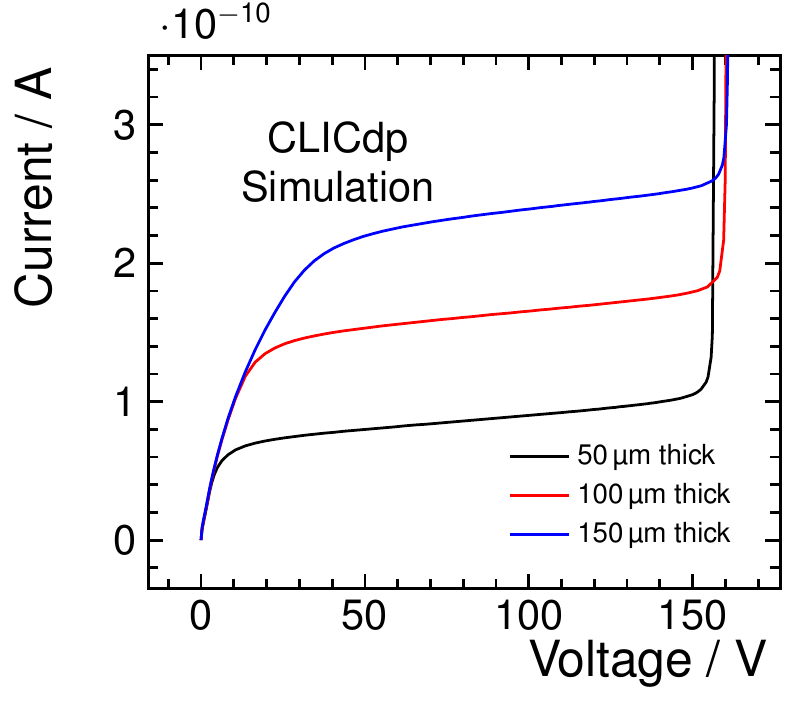}
  \caption{Simulated IV characteristics of the \SI{50}{\micron}, \SI{100}{\micron} and \SI{150}{\micron} thick active-edge sensors with grounded guard ring and \SI{55}{\micron} edge distance.}\label{fig:iv_thickness}
\end{minipage}
\end{figure*}

Due to the extension of the backside electrode to the sensor edge, the electric field in the region between the edge implant and the pixel implant can be high. This makes the sensor prone to early breakdowns. To extract the breakdown voltage, the voltage on the backside electrode is increased above the nominal operation point until a steep increase of the reverse current is observed. \cref{fig:iv_measurement} shows the measured IV characteristics of the full sensor for the investigated edge geometries, and \cref{fig:iv_50um} for two simulated pixels and the different simulated edge termination schemes for the \SI{50}{\micron} thick sensors. Above a certain electric field strength, avalanche ionisation is releasing plenty of free charge carriers in the silicon bulk and thus the breakdown is closely linked to the distribution of the electric field. Most relevant is the distance between the last pixel implant and the active edge.

In data, the devices without guard ring and edge distance of \SI{20}{\micron} and \SI{50}{\micron} show a similar breakdown around \SI{125}{\volt}, whereas in the device with floating guard ring and edge distance of \SI{50}{\micron}, the breakdown voltage is improved to about \SI{180}{\volt}. In the device with floating guard ring and narrow edge distance of \SI{23}{\micron}, the breakdown appears already around \SI{90}{\volt}. In that particular design, the guard ring is routed very close to the pixel implant, especially in the matrix corner, which is expected to degrade the breakdown behaviour.

In the device simulations, both sensors without guard ring show the breakdown of the junction around \SI{150}{\volt}, with slightly higher breakdown voltage of the wider edge layout, similar to the experimental results. The addition of a floating guard ring smoothens the voltage drop in the silicon bulk close to the surface and reduces the electric field strength at a given voltage, which results in a higher breakdown voltage. For the narrow edge layout, the breakdown voltage is significantly increased, up to \SI{240}{\volt}. The beneficial effect is smaller in the layout with \SI{50}{\micron} edge distance, which might be a result of the different distances between edge, guard ring and pixel implant. This indicates room for further optimisation of the various layouts, if needed.

The breakdown behaviour is qualitatively similar to the measured values, with the exception of the narrow device with floating guard ring. Experimentally, the breakdown has been observed already below \SI{100}{\volt}, which is not reproduced in simulation. However, exact agreement is not expected due to the simplifications taken in the two-dimensional simulation setup, e.g. neglecting the matrix corner pixels.

The potential gradient between the edge implant and the guard ring does not significantly depend on the sensor thickness, if the same edge layout is implemented. For that reason, the simulated breakdown in the \SI{50}{\micron}, \SI{100}{\micron} and \SI{150}{\micron} thick sensors with grounded guard ring is appearing around \SI{150}{\volt} for all thicknesses, as illustrated in \cref{fig:iv_thickness}.

\subsection{Charge collection and efficiency in \SI{50}{\micron} thick sensors}
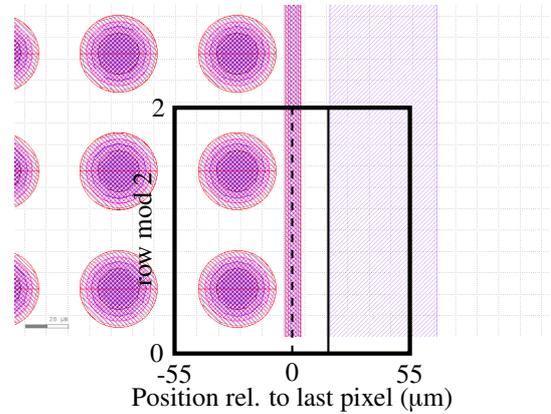
\begin{figure}[tbp]
  \centering
  \begin{tikzpicture}
    \node[anchor=south west,inner sep=0] (image) at (0,0) {
      \includegraphics[width=.9\linewidth]{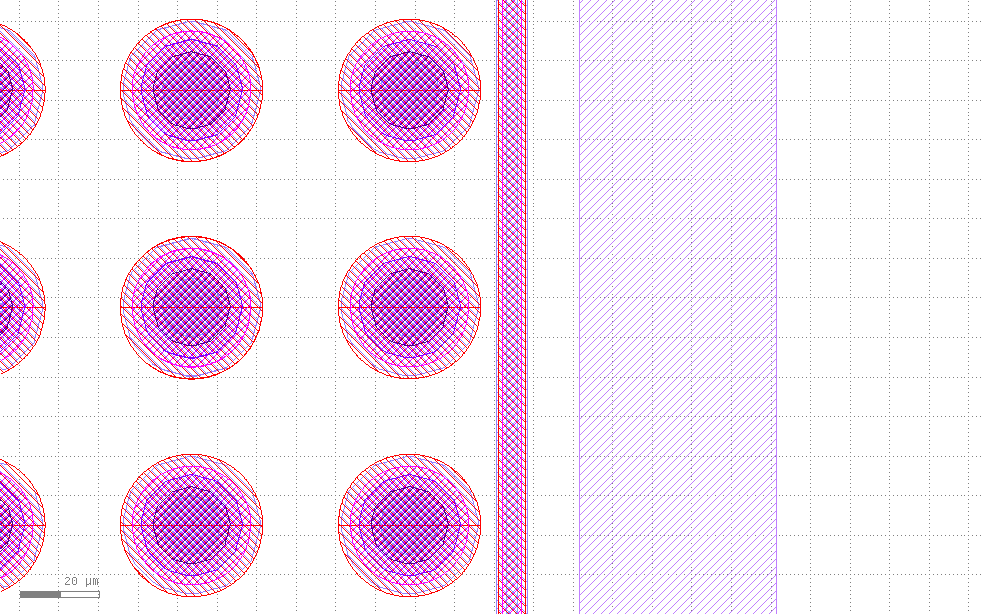}};
  \begin{scope}[x={(image.south east)},y={(image.north west)}]
    \draw[ultra thick] (0.3,-0.05) node[below]{-55} node[left]{0} rectangle ++(.44,.74) ++(0,-0.74) node[below]{55};
    \node[left] at (0.3,0.69){2};
    \node[anchor=south,rotate=90] at (0.275,0.325){row mod 2};
  \draw[thick, dashed](0.3,-0.05) ++(0.22,0) node[below]{0} -- +(0,0.74);
  \draw[thick](0.5875,-0.05) -- ++(0,0.74);
  \node[anchor=north] at (0.52,-0.125){Position rel. to last pixel (\si{\micron})};
    \end{scope}
  \end{tikzpicture}
  \caption{Definition of the coordinate system for the in-pixel resolved efficiency studies: For all 256 pixel rows, the last pixel column and the edge region is considered.}\label{fig:eff_definition}
\end{figure}

Exploiting the good pointing capabilities of the reference telescope, the collected signal and the hit efficiency of the active-edge region have been studied as a function of the track incident point. The efficiency has been mapped in two dimensions. For the last two columns, all 256 pixel rows are mapped to a \num{2x2} pixel plot (see \cref{fig:eff_definition} for definition of the coordinates).

\begin{figure*}[tbp]
  \begin{minipage}{\linewidth}
    \begin{subfigure}[T]{.24\linewidth}
      \begin{tikzpicture}
        \node[anchor=south west,inner sep=0] (image) at (0,0)
        {\includegraphics[page=5,width=\linewidth]{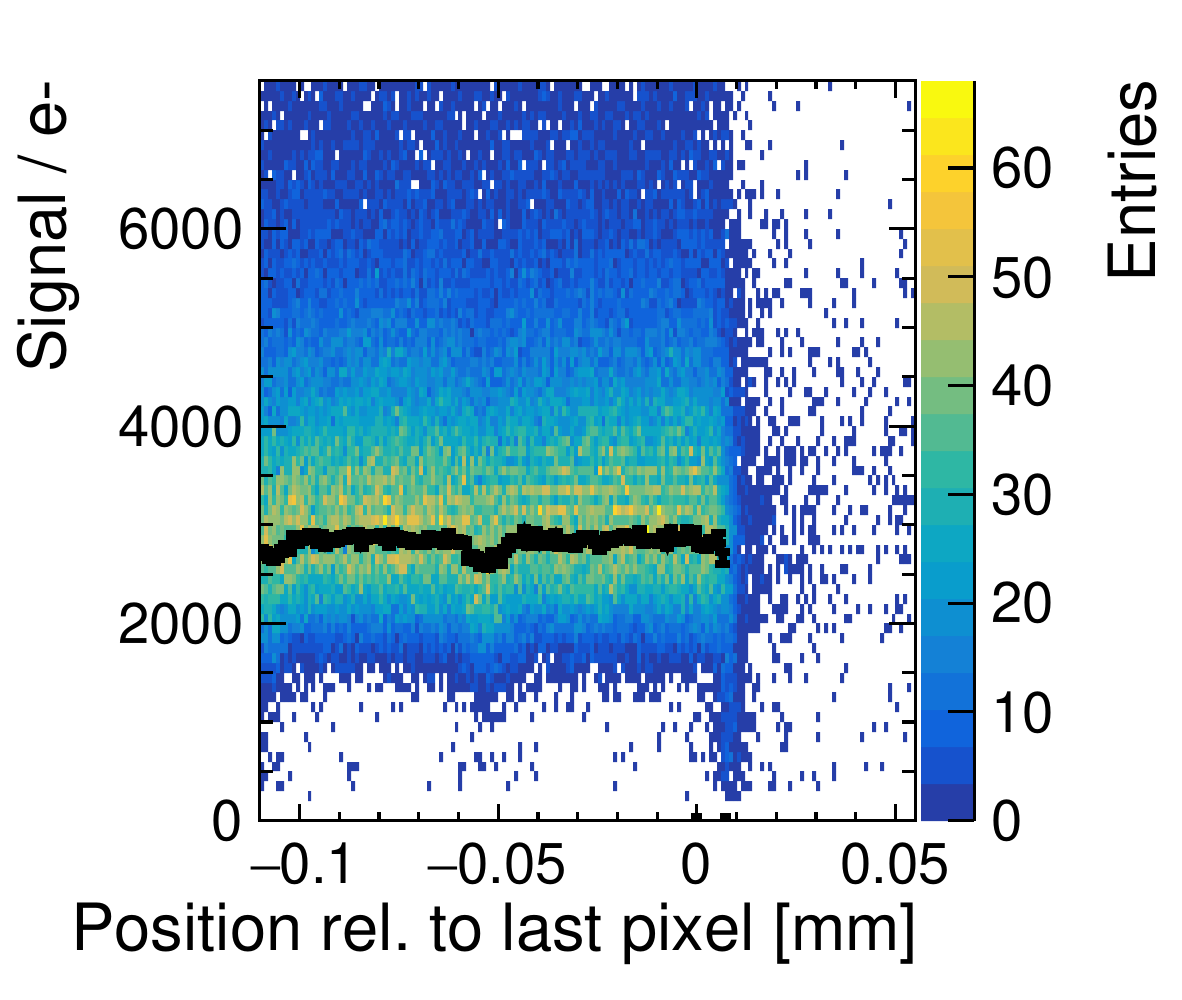}};
        \begin{scope}[x={(image.south east)},y={(image.north west)}]
            \draw[ultra thick](0.6125,0.18)--+(0,0.74);
           \node[anchor= west] at  (0.2,0.97){\textcolor{black}{\scriptsize\sffamily CLICdp}};
        \end{scope}
      \end{tikzpicture}
      \caption{No GR, \SI{20}{\micron}}\label{fig:signal_nogr}
    \end{subfigure}
    \begin{subfigure}[T]{.24\linewidth}
      \begin{tikzpicture}
        \node[anchor=south west,inner sep=0] (image) at (0,0)
        {\includegraphics[page=8,width=\linewidth]{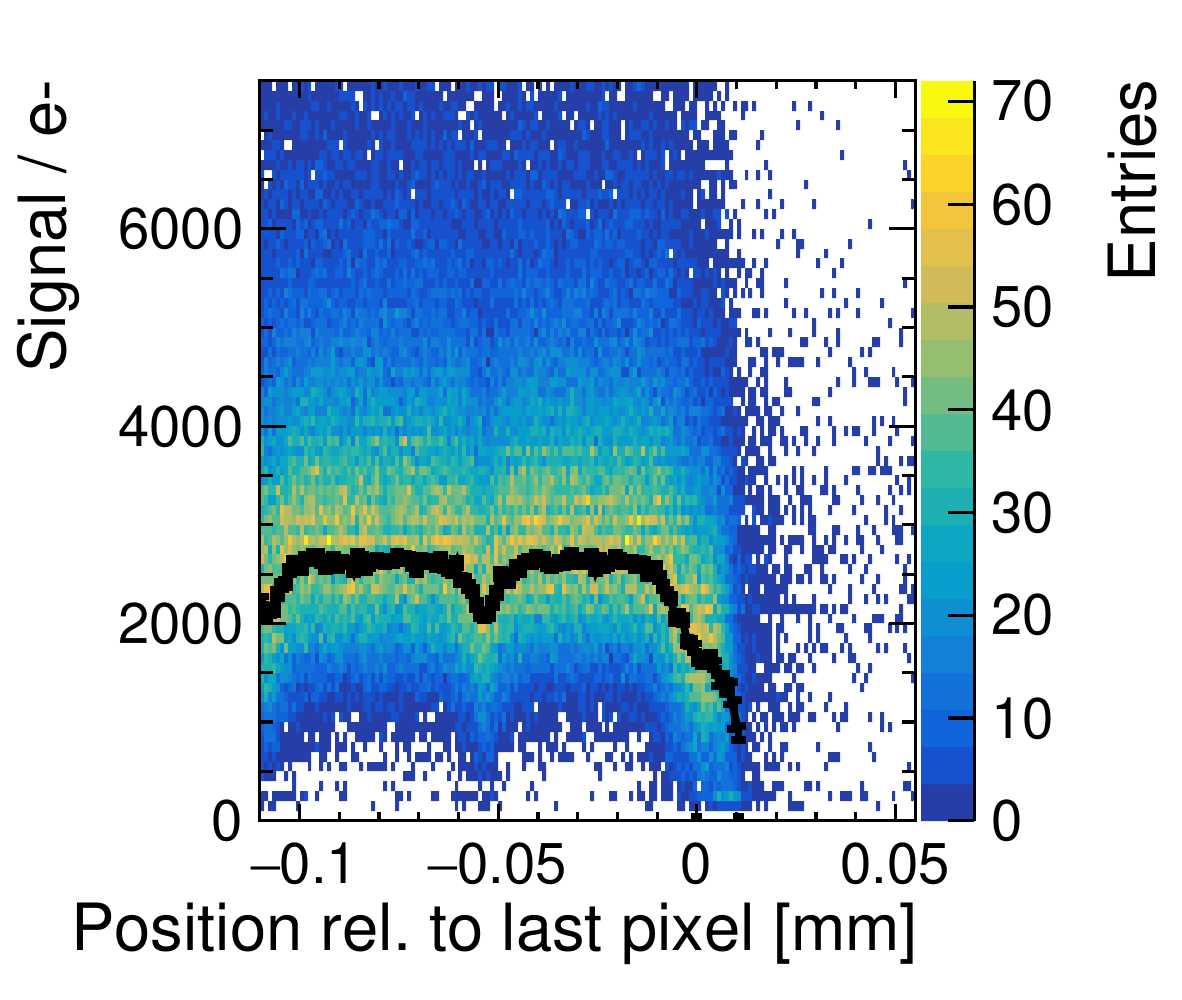}};
        \begin{scope}[x={(image.south east)},y={(image.north west)}]
            \draw[ultra thick](0.625,0.18)--+(0,0.74);
             \node[anchor= west] at  (0.2,0.97){\textcolor{black}{\scriptsize\sffamily CLICdp}};
        \end{scope}
      \end{tikzpicture}
      \caption{Floating GR, \SI{23}{\micron}}\label{fig:signal_floatgr}
    \end{subfigure}
    \begin{subfigure}[T]{.24\linewidth}
      \begin{tikzpicture}
        \node[anchor=south west,inner sep=0] (image) at (0,0)
        {\includegraphics[page=11,width=\linewidth]{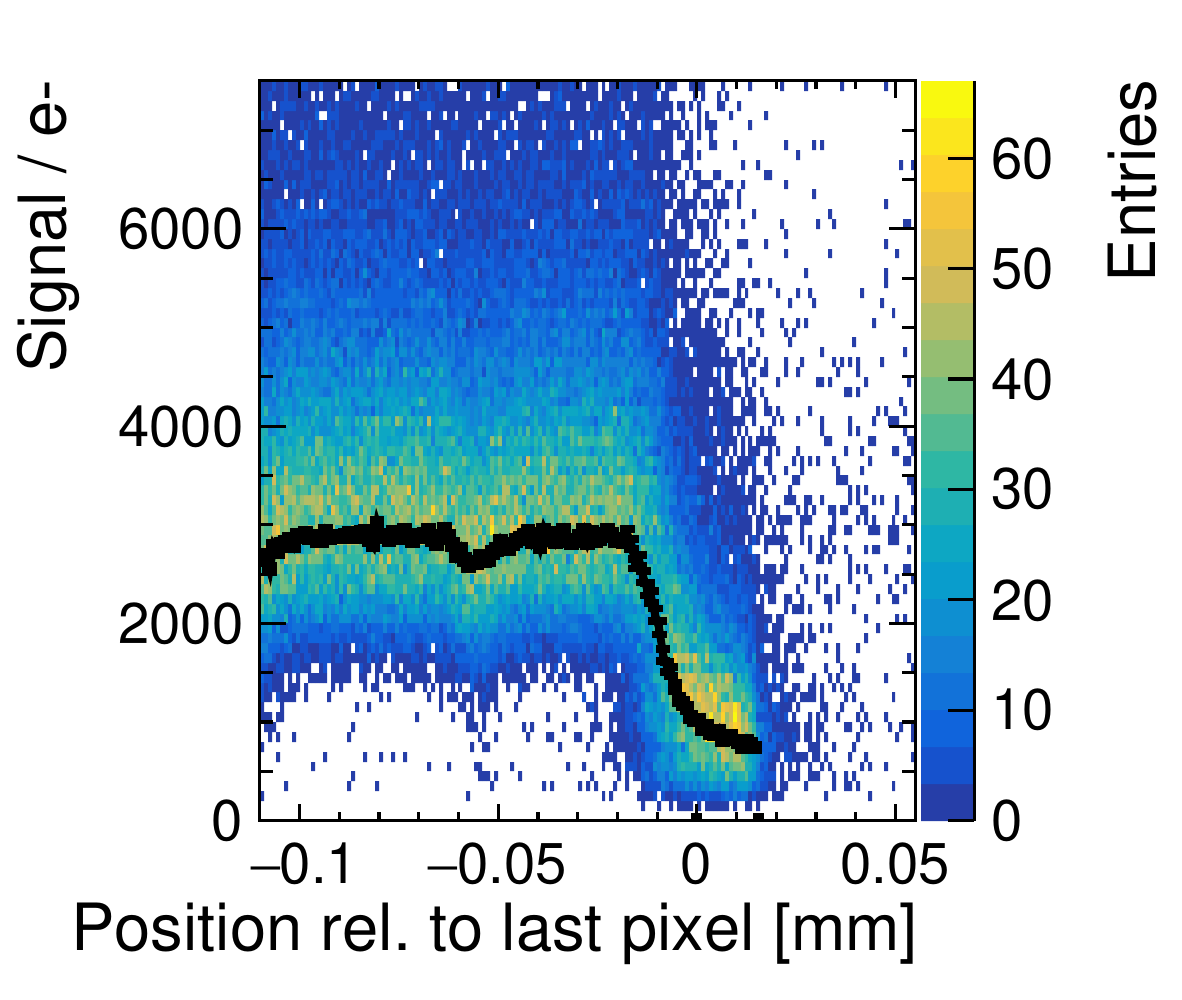}};
        \begin{scope}[x={(image.south east)},y={(image.north west)}]
            \draw[ultra thick](0.64,0.18)--+(0,0.74);
             \node[anchor= west] at  (0.2,0.97){\textcolor{black}{\scriptsize\sffamily CLICdp}};
        \end{scope}
      \end{tikzpicture}
      \caption{Grounded GR, \SI{28}{\micron}}\label{fig:signal_gndgr28}
    \end{subfigure}
    \begin{subfigure}[T]{.24\linewidth}
      \begin{tikzpicture}
        \node[anchor=south west,inner sep=0] (image) at (0,0)
        {\includegraphics[page=14,width=\linewidth]{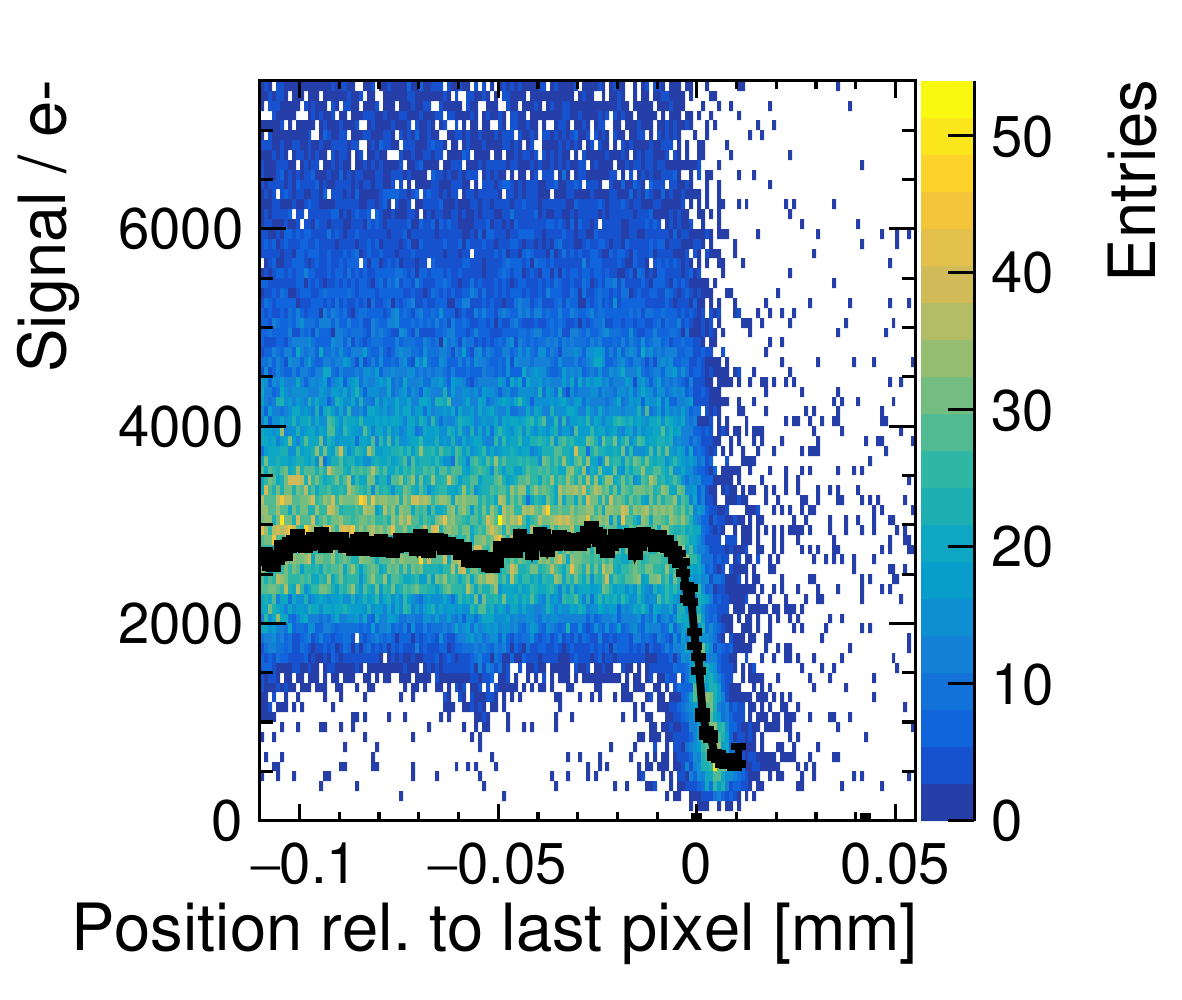}};
        \begin{scope}[x={(image.south east)},y={(image.north west)}]
            \draw[ultra thick](0.735,0.18)--+(0,0.74);
             \node[anchor= west] at  (0.2,0.97){\textcolor{black}{\scriptsize\sffamily CLICdp}};
        \end{scope}
      \end{tikzpicture}
      \caption{Grounded GR, \SI{55}{\micron}}\label{fig:signal_gndgr55}
      \end{subfigure}
    \caption{Experimentally obtained signal distribution close to the sensor edge for \SI{50}{\micron} thick sensors with different edge layout. 
    The vertical solid line indicates the physical edge of the sensor. The solid black line superimposed on the histogram indicates the most probable value of the cluster signal per x-bin.}\label{fig:signal_edge}
  \end{minipage}
  
  \begin{minipage}{\linewidth}
    \begin{subfigure}[T]{.24\linewidth}
      \begin{tikzpicture}
        \node[anchor=south west,inner sep=0] (image) at (0,0)
        {\includegraphics[page=3,width=\linewidth]{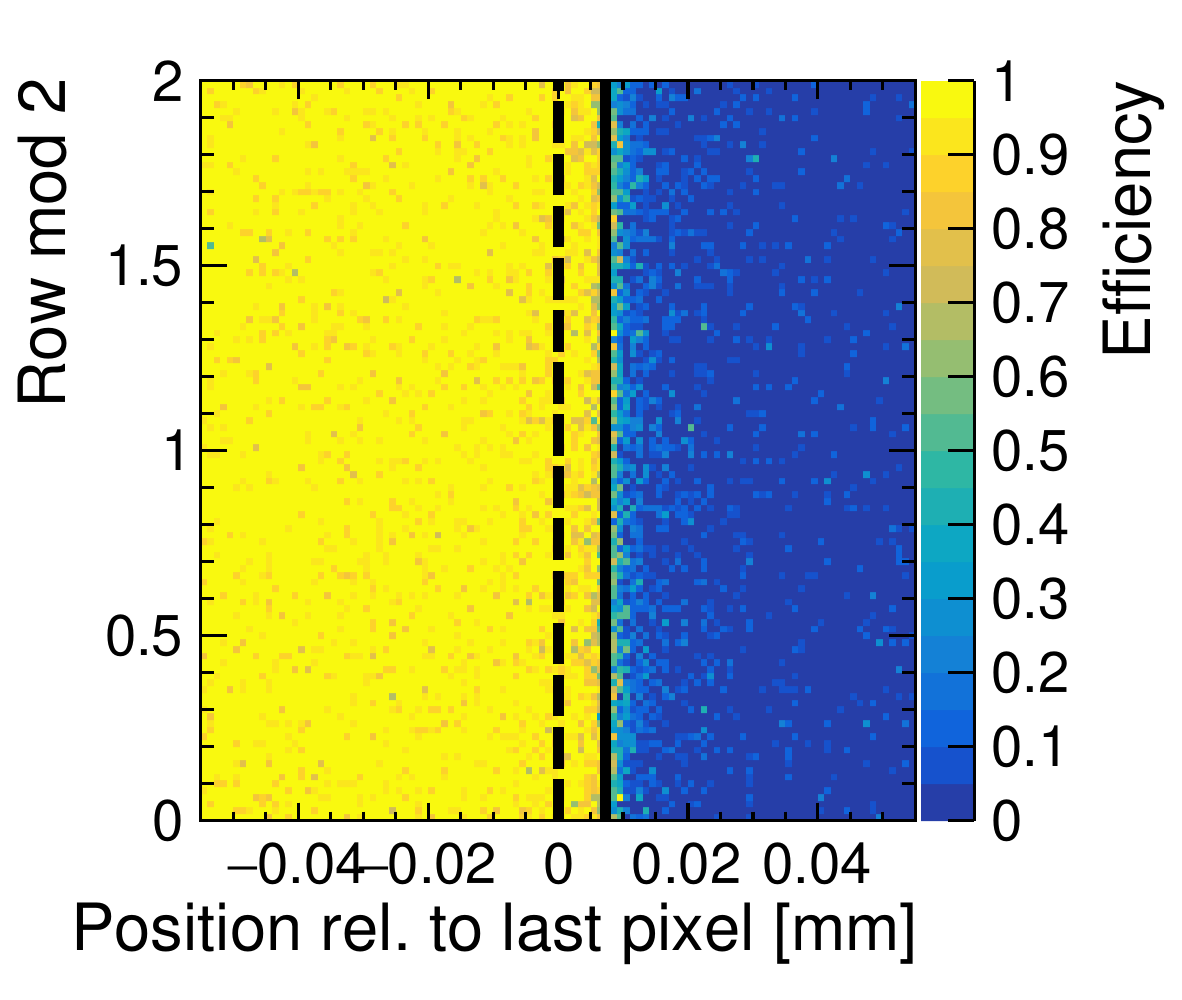}};
        \begin{scope}[x={(image.south east)},y={(image.north west)}]
          \node[anchor=south west] at (.2,.9){\scriptsize\sffamily CLICdp};
        \end{scope}
      \end{tikzpicture}
      \caption{No GR, \SI{20}{\micron}}\label{fig:eff_nogr}
    \end{subfigure}
      \begin{subfigure}[T]{.24\linewidth}
        \begin{tikzpicture}
          \node[anchor=south west,inner sep=0] (image) at (0,0)
          {\includegraphics[page=6,width=\linewidth]{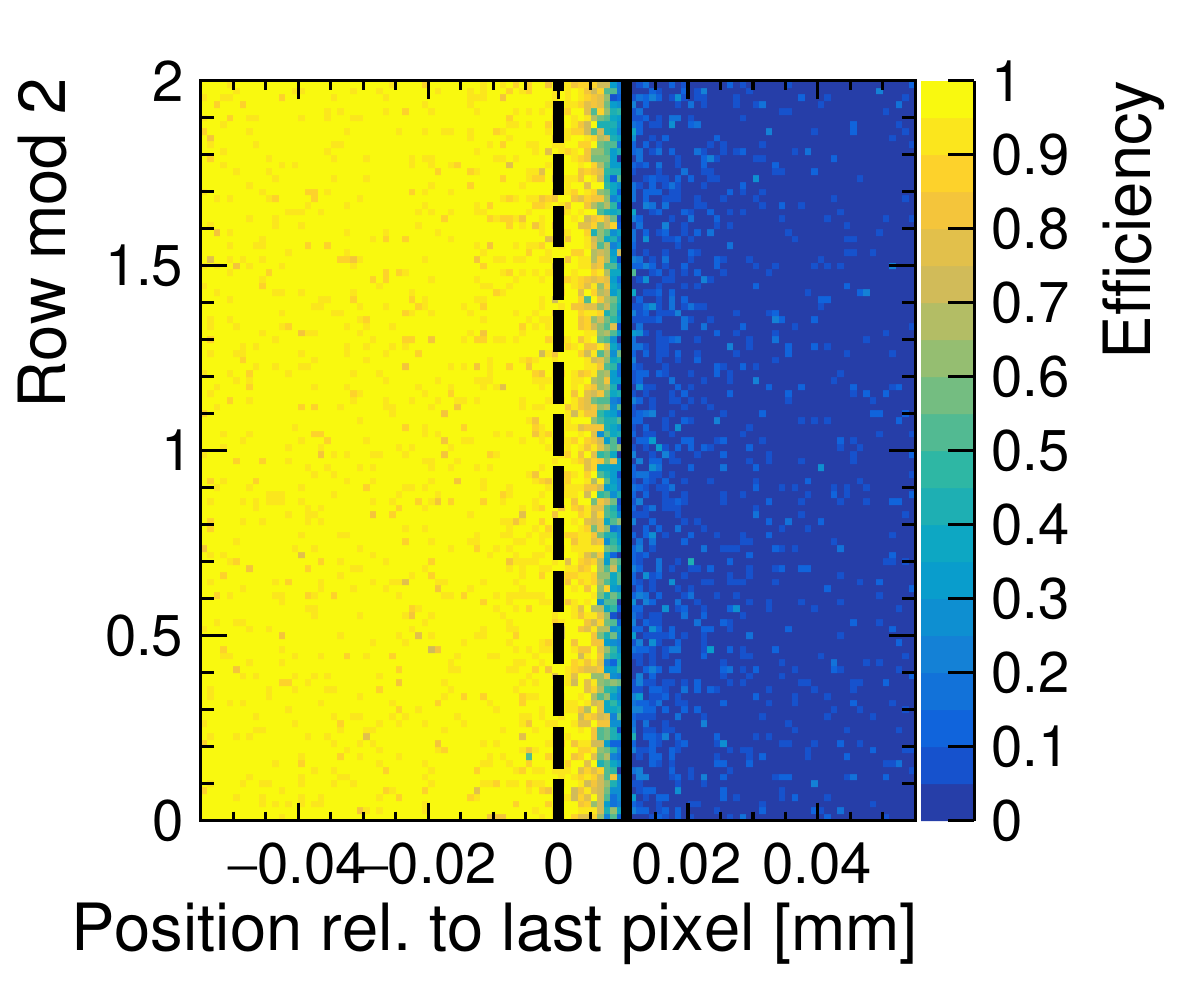}};
          \begin{scope}[x={(image.south east)},y={(image.north west)}]
            \node[anchor=south west] at (.2,.9){\scriptsize\sffamily CLICdp};
          \end{scope}
        \end{tikzpicture}
      \caption{Floating GR, \SI{23}{\micron}}\label{fig:eff_floatgr}
    \end{subfigure}    
      \begin{subfigure}[T]{.24\linewidth}
        \begin{tikzpicture}
          \node[anchor=south west,inner sep=0] (image) at (0,0)
          {\includegraphics[page=9,width=\linewidth]{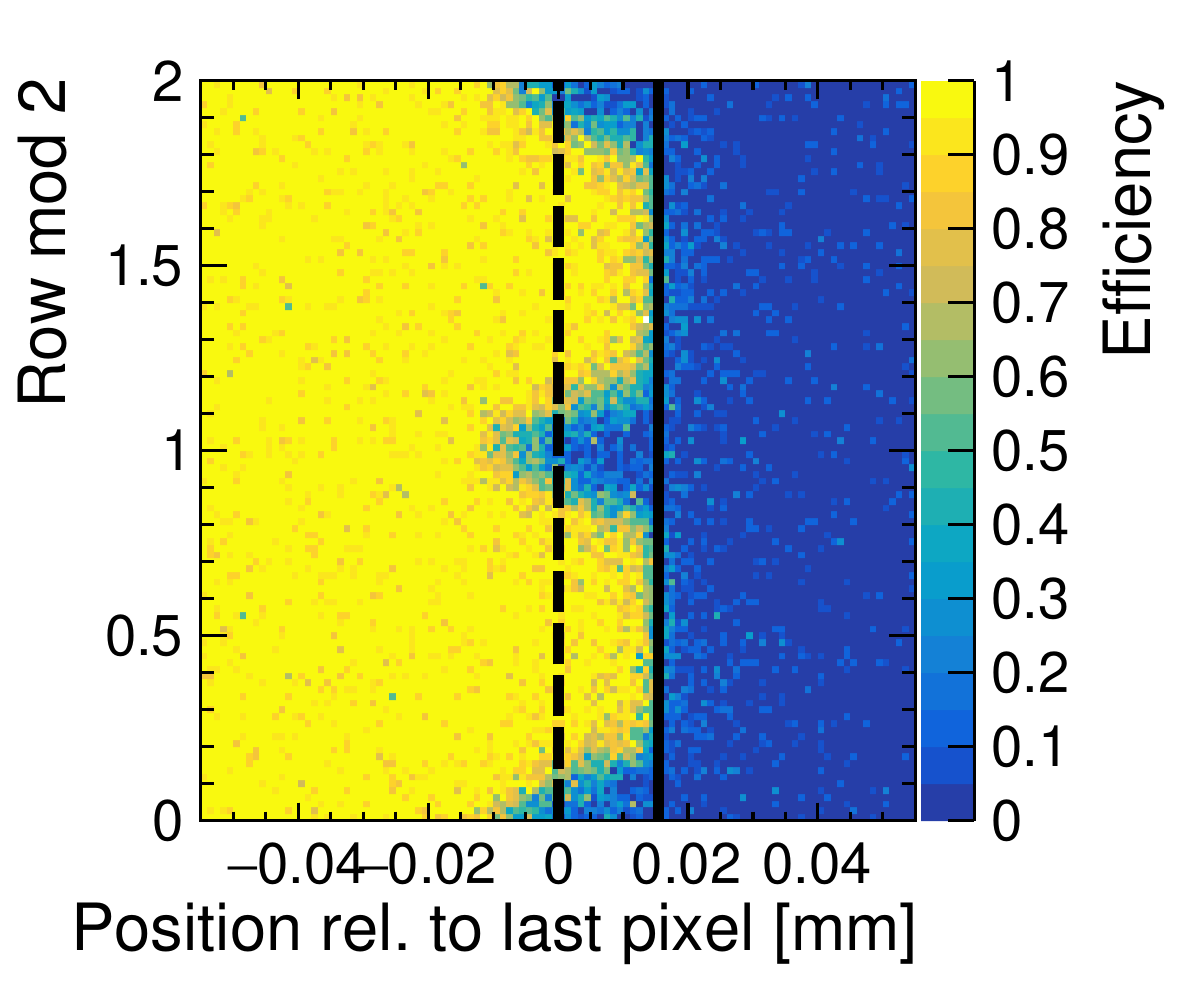}};
          \begin{scope}[x={(image.south east)},y={(image.north west)}]
            \node[anchor=south west] at (.2,.9){\scriptsize\sffamily CLICdp};
          \end{scope}
        \end{tikzpicture}
      \caption{Grounded GR, \SI{28}{\micron}}\label{fig:eff_gndgr28}
    \end{subfigure}
      \begin{subfigure}[T]{.24\linewidth}
        \begin{tikzpicture}
          \node[anchor=south west,inner sep=0] (image) at (0,0)
          {\includegraphics[page=12,width=\linewidth]{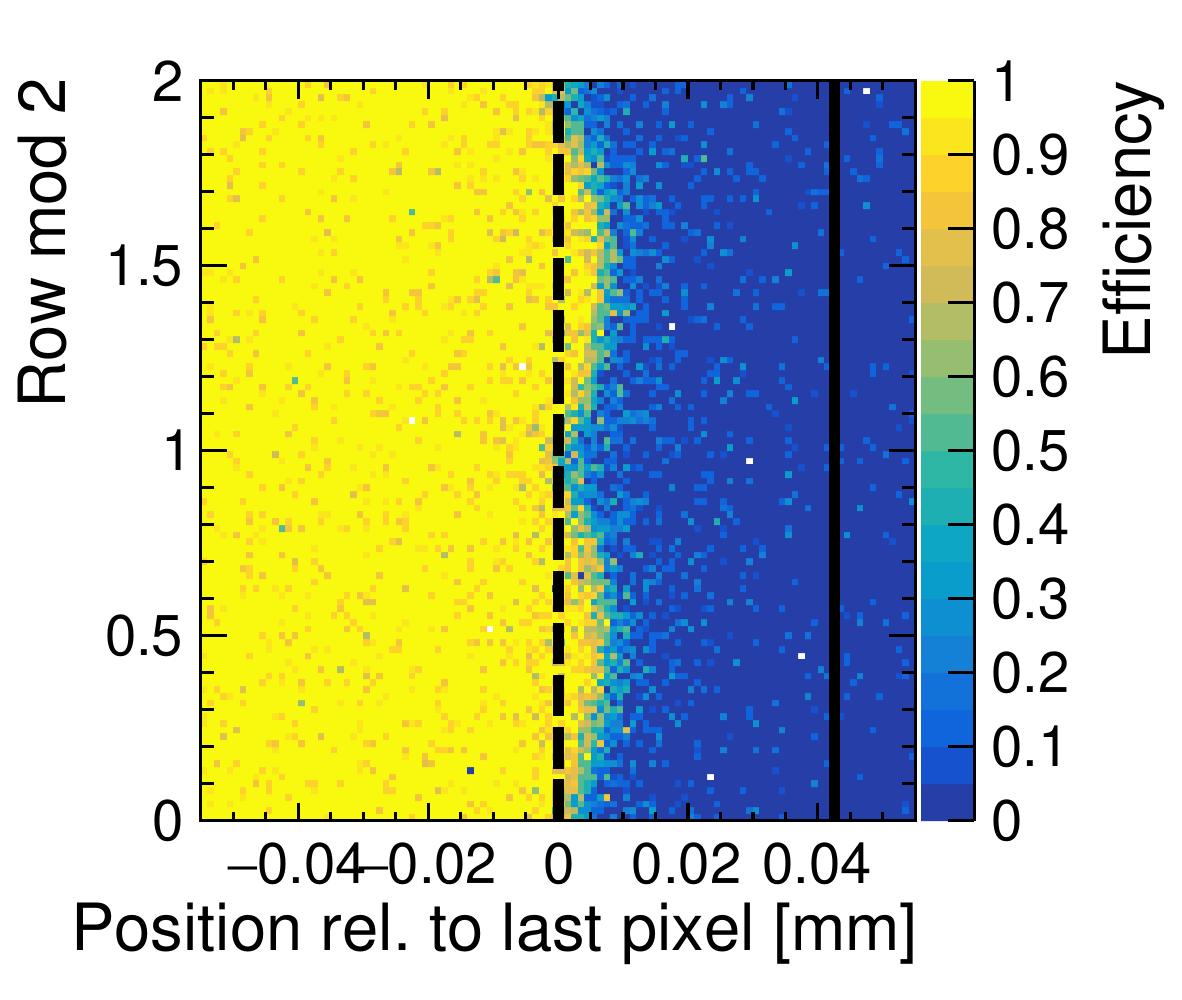}};
          \begin{scope}[x={(image.south east)},y={(image.north west)}]
            \node[anchor=south west] at (.2,.9){\scriptsize\sffamily CLICdp};
          \end{scope}
        \end{tikzpicture}
      \caption{Grounded GR, \SI{55}{\micron}}\label{fig:eff_gndgr55}
      \end{subfigure}
    \caption{Experimentally obtained detection efficiency close to the sensor edge for \SI{50}{\micron} thick sensors with different edge layout. 
     The dashed vertical line indicates the end of the regular pixel structure, the solid line indicates the physical edge of the sensor.
    }
    \label{fig:eff_edge}
  \end{minipage}
  
  \begin{minipage}{\linewidth}
    \vspace{.3cm}
    \begin{tikzpicture}
      \begin{groupplot}[group style = {group name=my plots,group size = 4 by 1,vertical sep=1.25cm ,horizontal sep=1.5cm},
        /pgf/number format/1000 sep={},
        axis line style={black},
        ticklabel style={black,font={\sansmath\sffamily\footnotesize}},
        label style={black,font={\sansmath\sffamily\footnotesize}},
        every tick/.style={black},
        xlabel = Position rel. to last pixel (\si{\micron}),
        ylabel= Signal (\Pem{}),
        ylabel style={at=(current axis.above origin), anchor=east,yshift=1.125cm},
        width=4cm,
        height=4cm,
        xmin=-60,xmax=40,
        ymin=0,
        ymax=3500,
        minor y tick num={4},
        minor x tick num={4},
        every axis/.append style={font=\footnotesize},
        legend style={at={(0.05,0.05)},anchor=south west,nodes=right,font=\tiny},
        ]

        \nextgroupplot
        \addplot[mark=none,thick,color=orange,/pgfplots/error bars/.cd, x dir=none,y dir=none, y explicit] plot[] table[x=Position, y expr=\thisrow{MPV}, y error expr=\thisrow{MPV_error}]{figures/W19_G7.dat}; 
        \addlegendentry{Data}; 
        \addplot[mark=o,only marks,thick,color=orange] plot[] table[x expr=51.5-\thisrow{hit_position}, y expr={\thisrow{sumcharge}}]{figures/noguard.sim.55}; 
        \addlegendentry{Sim}; 

        \nextgroupplot
        \addplot[mark=none,thick,color=orange,/pgfplots/error bars/.cd, x dir=none,y dir=none, y explicit] plot[] table[x=Position, y expr=\thisrow{MPV}, y error expr=\thisrow{MPV_error}]{figures/W19_F7.dat}; 
        \addlegendentry{Data}; 
        \addplot[mark=o,only marks,thick,color=orange] plot[] table[x expr=61.5-\thisrow{hit_position}, y expr={\thisrow{sumcharge}}]{figures/floatgr.sim.55}; 
        \addlegendentry{Sim};

        \nextgroupplot			
        \addplot[mark=none,thick,color=orange,/pgfplots/error bars/.cd, x dir=none,y dir=none, y explicit] plot[] table[x=Position, y expr=\thisrow{MPV}, y error expr=\thisrow{MPV_error}]{figures/W19_L8.dat}; 
        \addlegendentry{Data}; 
        \addplot[mark=o,only marks,thick,color=orange] plot[] table[x expr=61.5-\thisrow{hit_position}, y expr={\thisrow{sumcharge}}]{figures/grguard.sim.55}; 
        \addlegendentry{Sim}; 
        
        \nextgroupplot			
        
        \addplot[mark=none,thick,color=orange,/pgfplots/error bars/.cd, x dir=none,y dir=none, y explicit] plot[] table[x=Position, y expr=\thisrow{MPV}, y error expr=\thisrow{MPV_error}]{figures/W19_C7.dat}; 
        \addlegendentry{Data}; 
        \addplot[mark=o,only marks,thick,color=orange] plot[] table[x expr=91.5-\thisrow{hit_position}, y expr={\thisrow{sumcharge}}]{figures/grguard50_50thick.sim}; 
        \addlegendentry{Sim}; 
        \addplot[mark=otimes,only marks,thick,color=red] plot[] table[x expr=91.5-\thisrow{hit_position}, y expr={\thisrow{sumcharge}}]{figures/grguard50_50thick_nothresh.sim}; 

      \end{groupplot}
      \node[below = 12mm of my plots c1r1.south] {\small (a) No GR, \SI{20}{\micron}};
      \node[below = 12mm of my plots c2r1.south] {\small (b) Floating GR, \SI{23}{\micron}};
      \node[below = 12mm of my plots c3r1.south] {\small (c) Grounded GR, \SI{28}{\micron}};
      \node[below = 12mm of my plots c4r1.south] {\small (d) Grounded GR, \SI{55}{\micron}};
      
      \node[above = 2mm of my plots c1r1.north,anchor=east] {\textcolor{black}{\scriptsize CLICdp}};
      \node[above = 2mm of my plots c2r1.north,anchor=east]{\textcolor{black}{\scriptsize\sffamily CLICdp}};
      \node[above = 2mm of my plots c3r1.north,anchor=east]{\textcolor{black}{\scriptsize\sffamily CLICdp}};
      \node[above = 2mm of my plots c4r1.north,anchor=east]{\textcolor{black}{\scriptsize\sffamily CLICdp}};
      
    \end{tikzpicture}
    \caption{Comparison of the simulated signal near the sensor edge to the experimentally obtained most probable signal value for \SI{50}{\micron} thick sensors with different edge layout (cf. \cref{fig:signal_edge}). For the device with \SI{55}{\micron} wide edge and grounded guard ring, the simulated signal drops below the detection threshold of $\sim\SI{600}{\Pem{}}$ applied in the experimental studies. This is indicated by the style change in the simulated curve.}\label{fig:signal_comparison}
  \end{minipage}
\end{figure*}

The experimentally measured charge collection close to the sensor edge is summarised in \cref{fig:signal_edge}. As already illustrated in \cref{fig:sim_cross_section}, for better comparability with the 2d-simulations, only particle tracks which pass the sensor within the central \SI{40}{\percent{}} of the pixel cell area are considered.

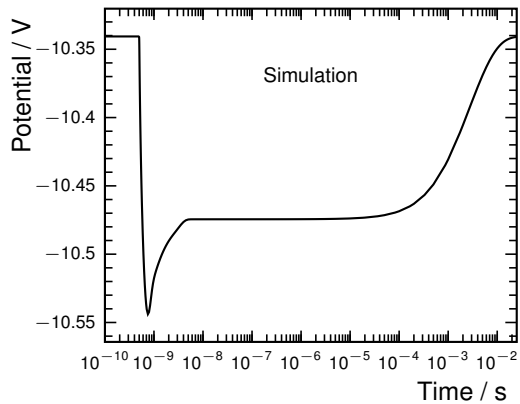
\begin{figure}[tbp]
\begin{tikzpicture}
  \begin{axis}[
    /pgf/number format/1000 sep={},
    axis line style={black,thick},
    ticklabel style={black,font={\sansmath\sffamily\scriptsize}},
    every tick/.style={black,semithick},
    xlabel= Time / \si{\second},
    ylabel = Potential / \si{\volt},
    ylabel near ticks,
    xlabel near ticks,
    label style={black,font={\sansmath\sffamily}},
    xlabel style={at={(rel axis cs:1,0)}, anchor=north east,yshift=-.4cm},
    ylabel style={at={(rel axis cs:0,1)}, anchor=south east,yshift=.85cm},
    width=7cm,
    height=6cm,
    xmin=1e-10,
    xmax=0.025,
    minor y tick num={4},
    minor x tick num={4},
    try min ticks=5,
    max space between ticks=25,
    xmode=log,
    every axis/.append style={font=\footnotesize},
    legend style={at={(0.95,0.05)},anchor=south east,nodes=right,font={\tiny}},
    ]
  
    \addplot[mark=none,thick,color=black] plot[] table[x expr={\thisrow{guardRingOuterVoltageX}*1}, y=guardRingOuterVoltageY]{figures/guardring.csv}; 

    \node at (rel axis cs:0.5,0.8) {\textcolor{black}{\sansmath\sffamily Simulation}};
  \end{axis}
\end{tikzpicture}
\caption{Electric potential on a floating guard ring after a particle hit close to the sensor edge, \SI{500}{\ps} after the start of the simulation. Relaxation to the equilibrium is achieved only after several \si{\milli\second}. Due to the short peaking time of the Timepix3 ASIC of \SI{30}{\ns}, charge stored on the guard ring (in this example about \SI{330}{\Pem{}}) is lost for detection.}\label{fig:transient_potential_gr}
\end{figure}

In devices without guard ring, the recorded signal is constant up to the physical edge of the sensor, as the pixel implants are the only available contacts towards which the charges can drift. This is illustrated by the signal distribution as a function of the track incident position close to the sensor edge, shown in \cref{fig:signal_nogr}. In devices with floating guard ring, a slight drop of the signal for particles passing close to the edge is observed, as shown in \cref{fig:signal_floatgr}.

A significant loss of signal to grounded guard rings is observed for tracks passing the sensor after the last pixel, as visible in \cref{fig:signal_gndgr28,fig:signal_gndgr55}.

As already deduced from the signal distributions in \cref{fig:signal_edge}, the device without guard and the device with floating guard ring are fully efficient up to the physical edge of the silicon, whereas for both sensors with grounded guard ring a significant loss of efficiency can be observed, as shown in \cref{fig:eff_edge}. This can be attributed to the fact that the grounded guard ring and the last pixel implant are competing in collecting the ionisation charge. If the guard ring is the closest implant to the particle hit, most of the signal is collected there and is lost for detection. This is especially the case between two pixel rows.

For a comparison of the collected charge, a simulated particle hit has been scanned along the sensor. For each position, the total charge recorded by both pixels is summed up. However in the simulation, no per-pixel threshold is applied to the charge signal, and Landau fluctuations of the signal deposition and delta electrons are not taken into account.
The trends observed in the experimental data are reproduced by the simulation (see \cref{fig:signal_comparison}). The slight fluctuations on the simulation results are attributed to meshing effects close to the guard ring structure. For the device without guard ring, no signal drop at the sensor edge is observed, whereas the slight loss of signal to the floating guard ring is reproduced by the simulation model. For the two grounded guard ring options, a strong loss of signal is observed. In the wider edge configuration, the signal drops below the detection threshold of 600 to 1000 electrons, that has been set for the experimental studies, explaining the observed inefficiency of the device.

The signal drop in devices with floating guard ring can be explained by the capacitive coupling of the guard ring to the surrounding implants. Since the guard ring is not kept at a fixed potential, charge can be collected by the implant in the non-equilibrium state after a particle hit. In this particular case, the stored charge is about \SI{5.4e-17}{\coulomb} or \SI{330}{\Pem{}}. This additional charge on the guard ring implant results in a shift of the electrostatic potential of the implant due to its capacitance to the surrounding contacts, as illustrated in \cref{fig:transient_potential_gr} for a transient simulation of \SI{20}{\ms}. With time, the device returns to equilibrium, and the charge gets collected in the readout implants. The time needed to relax to the static state is longer than the \SI{30}{\ns} peaking time of the Timepix3 front-end, and thus charge stored on the guard ring does not contribute to the signal. In the simplified two-dimensional simulation, this effect can only be studied qualitatively, as the capacitance of the several centimetre long guard ring is not modelled accurately.

\subsection{Dependence of the edge efficiency on the sensor thickness}
\begin{figure*}[tbp]
  \begin{minipage}{\linewidth}
    \begin{subfigure}[T]{0.32\linewidth}
      \begin{tikzpicture}
        \node[anchor=south west,inner sep=0] (image) at (0,0)
        {\includegraphics[page=14,width=\linewidth]{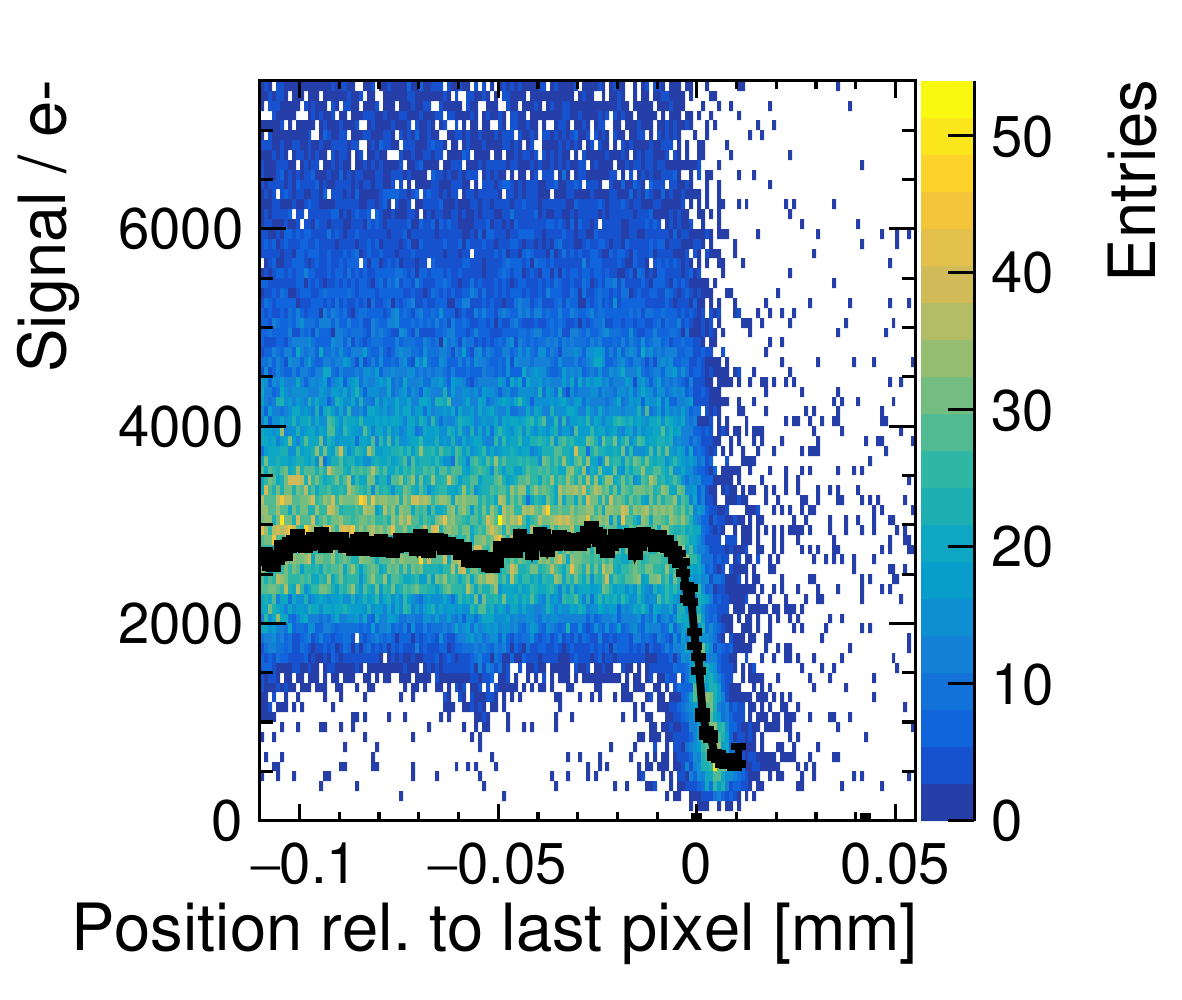}};
        \begin{scope}[x={(image.south east)},y={(image.north west)}]
             \draw[ultra thick](0.735,0.18)--+(0,0.74);
          \node[anchor=south west] at (.2,.9){\scriptsize\sffamily CLICdp};
        \end{scope}
      \end{tikzpicture}
      \caption{\SI{50}{\micron} thick}\label{fig:signal_50um}
    \end{subfigure}
    \begin{subfigure}[T]{0.32\linewidth}
      \begin{tikzpicture}
        \node[anchor=south west,inner sep=0] (image) at (0,0)
        {\includegraphics[page=17,width=\linewidth]{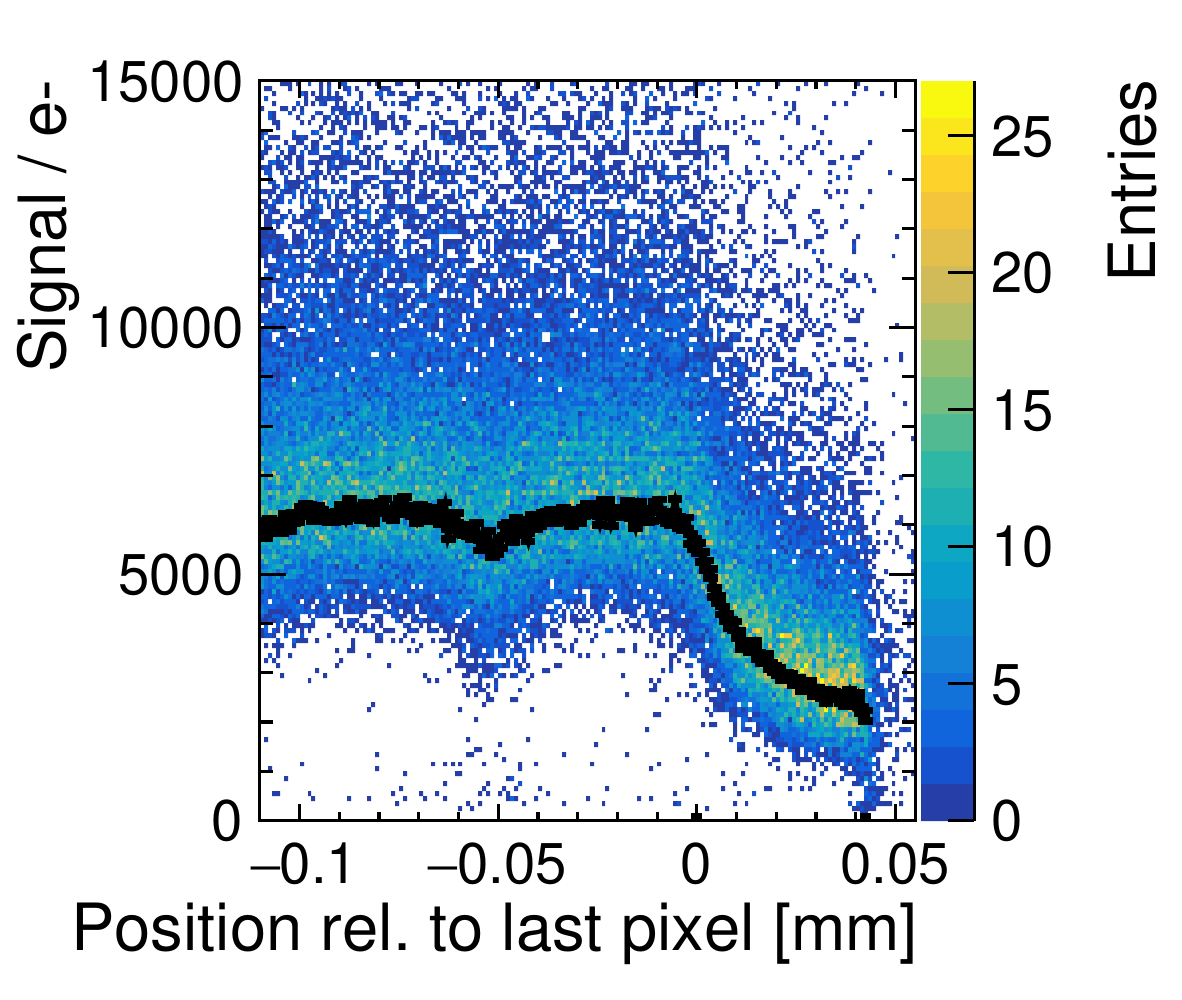}};
        \begin{scope}[x={(image.south east)},y={(image.north west)}]
          \draw[ultra thick](0.735,0.18)--+(0,0.74);
          \node[anchor=south west] at (.2,.9){\scriptsize\sffamily CLICdp};
        \end{scope}
      \end{tikzpicture}
      \caption{\SI{100}{\micron} thick}\label{fig:signal_100um}
    \end{subfigure}    
    \begin{subfigure}[T]{0.32\linewidth}
      \begin{tikzpicture}
        \node[anchor=south west,inner sep=0] (image) at (0,0)
        {\includegraphics[page=20,width=\linewidth]{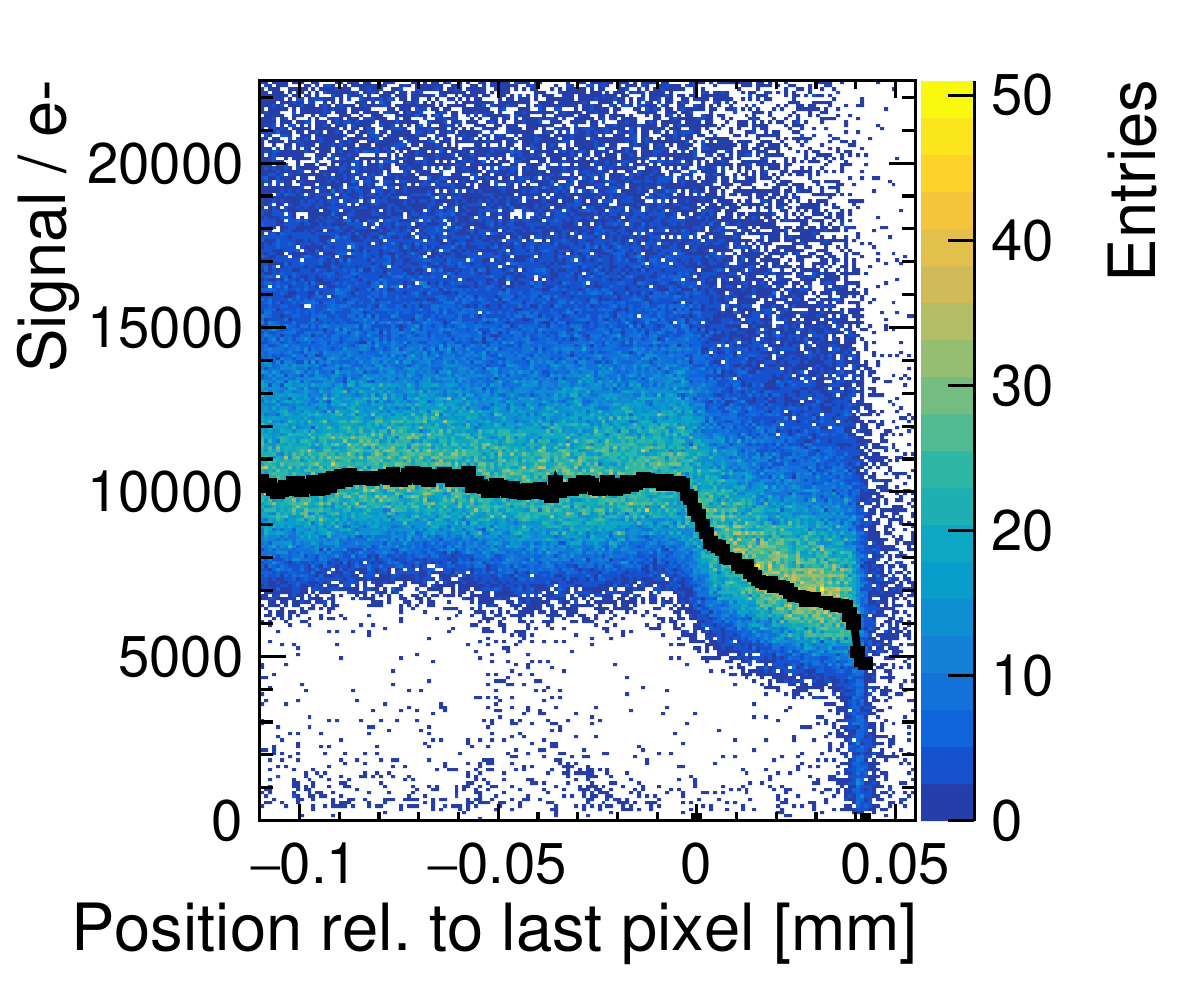}};
        \begin{scope}[x={(image.south east)},y={(image.north west)}]
          \draw[ultra thick](0.735,0.18)--+(0,0.74);
          \node[anchor=south west] at (.2,.9){\scriptsize\sffamily CLICdp};
        \end{scope}
      \end{tikzpicture}
      \caption{\SI{150}{\micron} thick}\label{fig:signal_150um}
    \end{subfigure}
    \caption{Signal distribution close to the sensor edge for \SI{50}{\micron}, \SI{100}{\micron} and \SI{150}{\micron} thick sensors with grounded guard ring and \SI{55}{\micron} edge distance. 
    The vertical solid line indicates the physical edge of the sensor. The solid black line superimposed on the histogram indicates the most probable value of the cluster signal per x-bin.}\label{fig:signal_edge_thickness}
  \end{minipage}
  
  \begin{minipage}{\linewidth}
    \begin{subfigure}[T]{0.32\linewidth}
      \begin{tikzpicture}
        \node[anchor=south west,inner sep=0] (image) at (0,0)
        {\includegraphics[page=12,width=\linewidth]{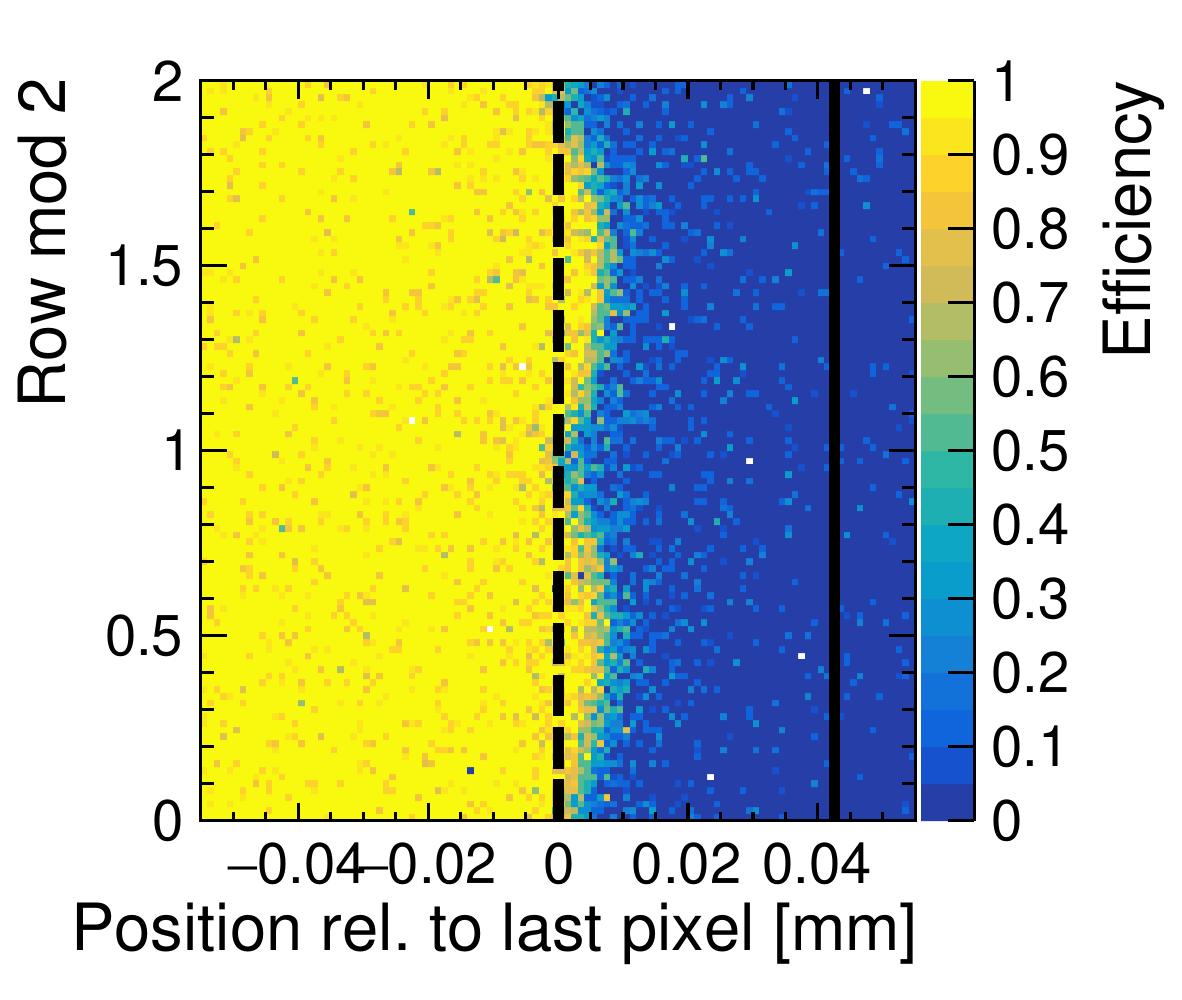}};
        \begin{scope}[x={(image.south east)},y={(image.north west)}]
          \node[anchor=south west] at (.2,.9){\scriptsize\sffamily CLICdp};
        \end{scope}
      \end{tikzpicture}
      \caption{\SI{50}{\micron} thick}\label{fig:eff_50um}
    \end{subfigure}
    \begin{subfigure}[T]{0.32\linewidth}
      \begin{tikzpicture}
        \node[anchor=south west,inner sep=0] (image) at (0,0)
        {\includegraphics[page=15,width=\linewidth]{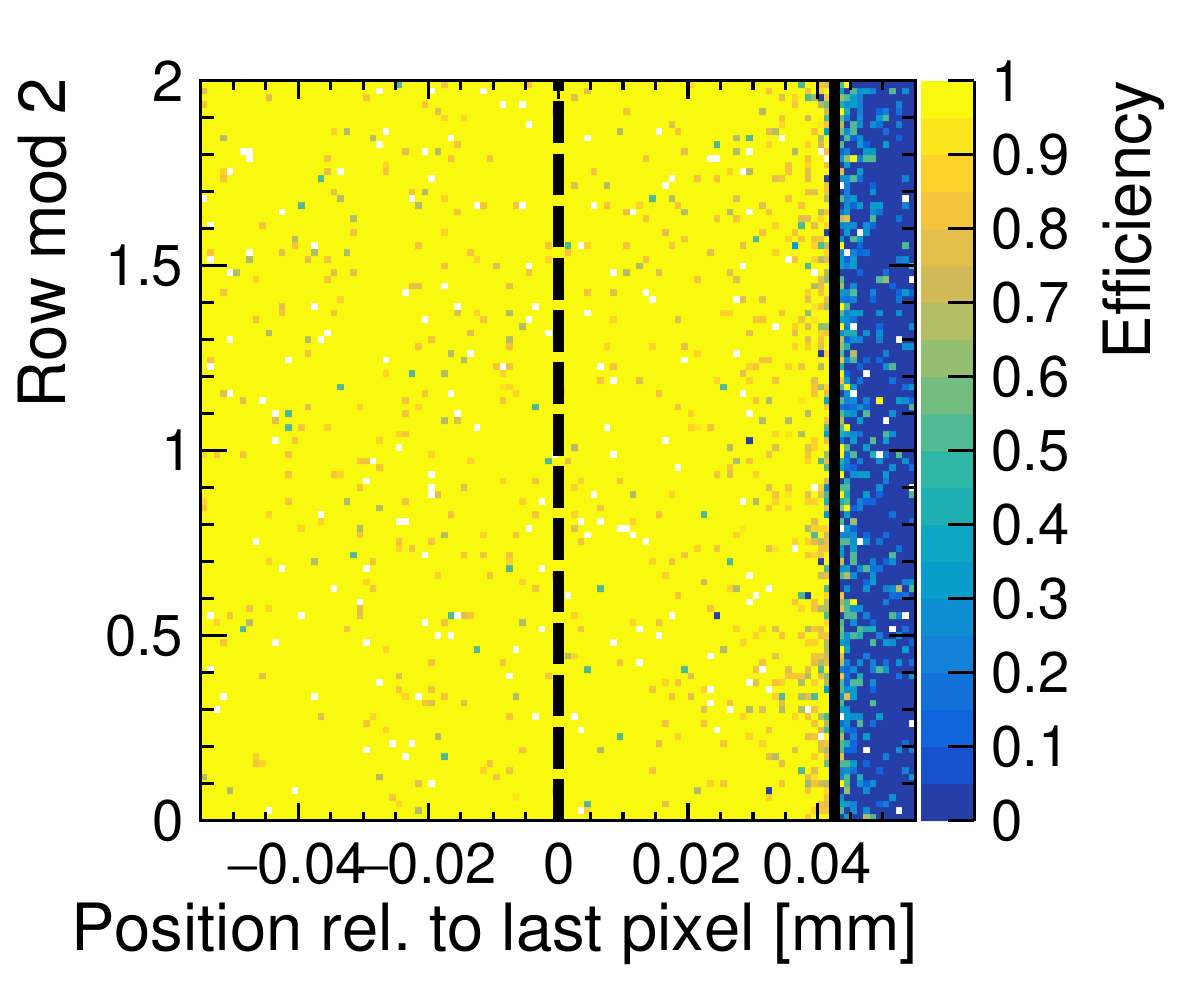}};
        \begin{scope}[x={(image.south east)},y={(image.north west)}]
          \node[anchor=south west] at (.2,.9){\scriptsize\sffamily CLICdp};
        \end{scope}
      \end{tikzpicture}
      \caption{\SI{100}{\micron} thick}\label{fig:eff_100um}
    \end{subfigure}    
    \begin{subfigure}[T]{0.32\linewidth}
      \begin{tikzpicture}
        \node[anchor=south west,inner sep=0] (image) at (0,0)
        {\includegraphics[page=18,width=\linewidth]{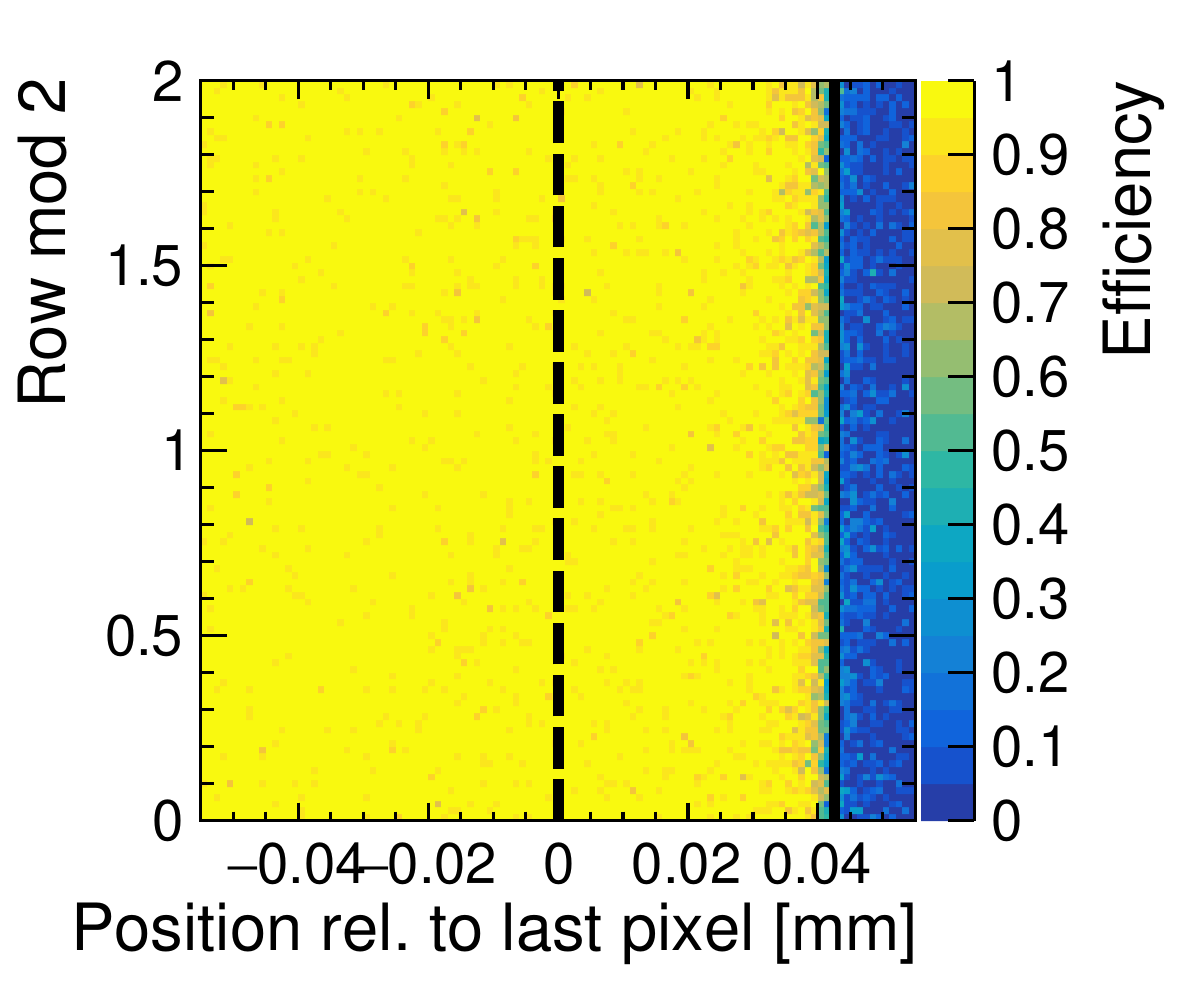}};
        \begin{scope}[x={(image.south east)},y={(image.north west)}]
          \node[anchor=south west] at (.2,.9){\scriptsize\sffamily CLICdp};
        \end{scope}
      \end{tikzpicture}
      \caption{\SI{150}{\micron} thick}\label{fig:eff_150um}
    \end{subfigure}
    \caption{Experimentally obtained efficiency map in \SI{50}{\micron}, \SI{100}{\micron} and \SI{150}{\micron} thick active-edge sensors with grounded guard ring and \SI{55}{\micron} edge distance. 
    The dashed vertical line indicates the end of the regular pixel structure, the solid line indicates the physical edge of the sensor.}\label{fig:efficiency_edge_thickness}
  \end{minipage}
  
  \begin{minipage}{\linewidth}
    \begin{tikzpicture}
      \begin{groupplot}[group style = {group name=my plots,group size = 3 by 1,vertical sep=1.5cm ,horizontal sep=2cm},
        /pgf/number format/1000 sep={},
        scaled y ticks = false,
        axis line style={black, thick},
        ticklabel style={black,font={\sansmath\sffamily\small}},
        label style={black,font={\sansmath\sffamily\small}},
        every tick/.style={black, thick},
        xlabel = Position rel. to last pixel (\si{\micron}),
        ylabel= Signal / \Pem{},
        ylabel style={at=(current axis.above origin), anchor=east,yshift=1.85cm},
        width=5cm,
        height=4.5cm,
        xmin=-55,xmax=40,
        ymin=0,
        minor y tick num={4},
        minor x tick num={4},
        every axis/.append style={font=\footnotesize},
        legend style={at={(0.05,0.05)},anchor=south west,nodes=right,font=\footnotesize},
        ]

        \nextgroupplot
        \addplot[mark=none,thick,color=orange,/pgfplots/error bars/.cd, x dir=none,y dir=none, y explicit] plot[] table[x=Position, y expr=\thisrow{MPV}, y error expr=\thisrow{MPV_error}]{figures/W19_C7.dat}; 
        \addlegendentry{Data}; 
        \addplot[mark=o,only marks,thick,color=orange] plot[] table[x expr=91.5-\thisrow{hit_position}, y expr={\thisrow{sumcharge}}]{figures/grguard50_50thick.sim}; 
        \addlegendentry{Sim}; 
        \addplot[mark=otimes,only marks,thick,color=red] plot[] table[x expr=91.5-\thisrow{hit_position}, y expr={\thisrow{sumcharge}}]{figures/grguard50_50thick_nothresh.sim}; 

        \nextgroupplot
        \addplot[mark=none,thick,color=orange,/pgfplots/error bars/.cd, x dir=none,y dir=none, y explicit] plot[] table[x=Position, y expr=\thisrow{MPV}, y error expr=\thisrow{MPV_error}]{figures/W5_E2.dat}; 
        \addlegendentry{\footnotesize Data}; 
        \addplot[mark=o,only marks,thick,color=orange] plot[] table[x expr=91.5-\thisrow{hit_position}, y expr={\thisrow{sumcharge}}]{figures/grguard50_100thick.sim}; 
        \addlegendentry{\footnotesize Sim};
        
        \nextgroupplot

        \addplot[mark=none,thick,color=orange,/pgfplots/error bars/.cd, x dir=none,y dir=none, y explicit] plot[] table[x=Position, y expr=\thisrow{MPV}, y error expr=\thisrow{MPV_error}]{figures/W5_F1.dat}; 
        \addlegendentry{\footnotesize Data}; 
        
        \addplot[mark=o,only marks,thick,color=orange] plot[] table[x expr=91.5-\thisrow{hit_position}, y expr={\thisrow{sumcharge}}]{figures/grguard50_150thick.sim}; 
        \addlegendentry{\footnotesize Sim};

      \end{groupplot}

      \node[below = 12mm of my plots c1r1.south] {\small (a) \SI{50}{\micron} thick};
      \node[below = 12mm of my plots c2r1.south] {\small (b) \SI{100}{\micron} thick};
      \node[below = 12mm of my plots c3r1.south] {\small (c) \SI{150}{\micron} thick};
      
      \node[above = 2mm of my plots c1r1.north,anchor=east] {\textcolor{black}{\scriptsize\sffamily CLICdp}};
      \node[above = 2mm of my plots c2r1.north,anchor=east]{\textcolor{black}{\scriptsize\sffamily CLICdp}};
      \node[above = 2mm of my plots c3r1.north,anchor=east]{\textcolor{black}{\scriptsize\sffamily CLICdp}};
      
    \end{tikzpicture}
    \caption{Comparison of the simulated signal distribution close to the edge to the experimentally obtained results in \SI{50}{\micron} (at \SI{-15}{\volt}), \SI{100}{\micron} (at \SI{-20}{\volt}) and \SI{150}{\micron} (at \SI{-30}{\volt}) thick active-edge sensors with grounded guard ring and \SI{55}{\micron} edge distance (cf. \cref{fig:signal_edge_thickness}). For the \SI{50}{\micron} thick device, the simulated signal drops below the detection threshold of $\sim\SI{600}{\Pem{}}$ applied in the experimental studies. This is indicated by the style change in the simulated curve.}\label{fig:signal_thickness}
  \end{minipage}
\end{figure*}

The performance of the active-edge devices depends on the sensor thickness, as illustrated in \cref{fig:signal_edge_thickness,fig:efficiency_edge_thickness,fig:signal_thickness}. For thin sensors, the charge loss to grounded guard rings can be significant and can lead to inefficient operation of the sensor. The grounded guard ring layout with \SI{55}{\micron} edge distance has been investigated experimentally on \SI{50}{\micron}, \SI{100}{\micron} and \SI{150}{\micron} thick sensors.
As in the \SI{50}{\micron} thin sensors, also in the thicker sensors a substantial charge loss to the guard ring is observed. This is illustrated in \cref{fig:signal_edge_thickness}, where the signal distribution close to the sensor edge for a \SI{50}{\micron}, \SI{100}{\micron} and \SI{150}{\micron} thick sensor is compared.

Due to the overall larger amount of ionisation charge in the thicker devices, the signal does not drop below the detection threshold. For that reason, particles passing the sensor close to its physical edge can still be detected efficiently. This is shown in the two-dimensional efficiency maps in \cref{fig:efficiency_edge_thickness}.

The qualitative agreement between the simulated and the experimentally obtained signal distribution is shown in \cref{fig:signal_thickness}.

\begin{figure}[tbp]
  \begin{subfigure}[b]{.31\linewidth}
    \begin{tikzpicture}
      \node[anchor=south west,inner sep=0] (image) at (0,0)
      {\includegraphics[width=.9\textwidth,clip=true,trim=2.5cm 5cm 13cm 3cm]{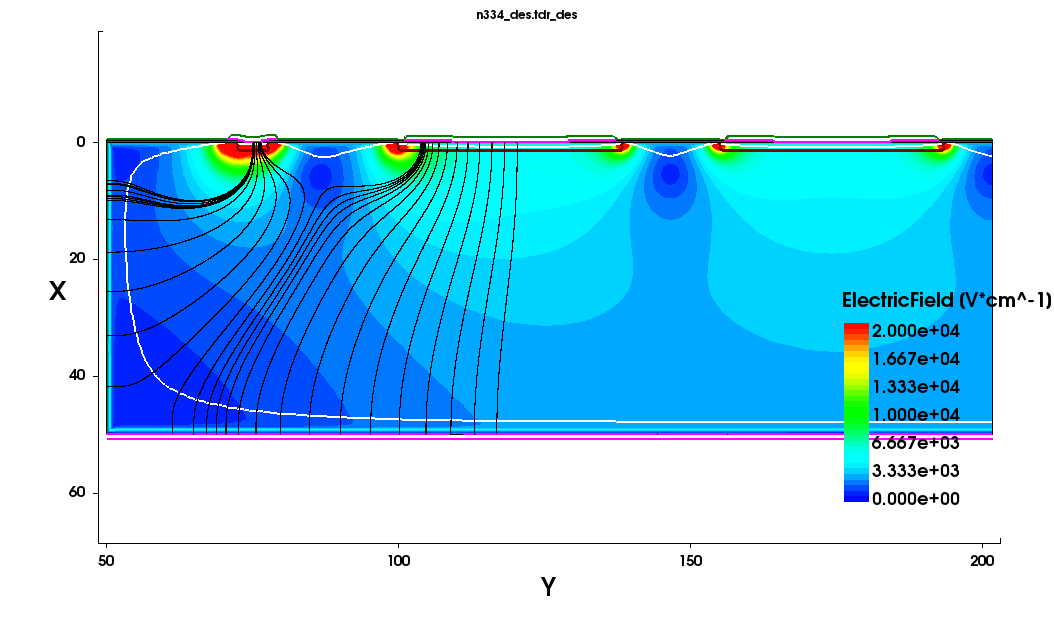}};
      \begin{scope}[x={(image.south east)},y={(image.north west)}]
      \end{scope}
    \end{tikzpicture}
    \caption{\SI{50}{\micron} thick}\label{fig:field_50um}
  \end{subfigure}
  \begin{subfigure}[b]{.31\linewidth}
    \begin{tikzpicture}
      \node[anchor=south west,inner sep=0] (image) at (0,0)
      {\includegraphics[width=.9\textwidth,clip=true,trim=4cm 2cm 12cm 0cm]{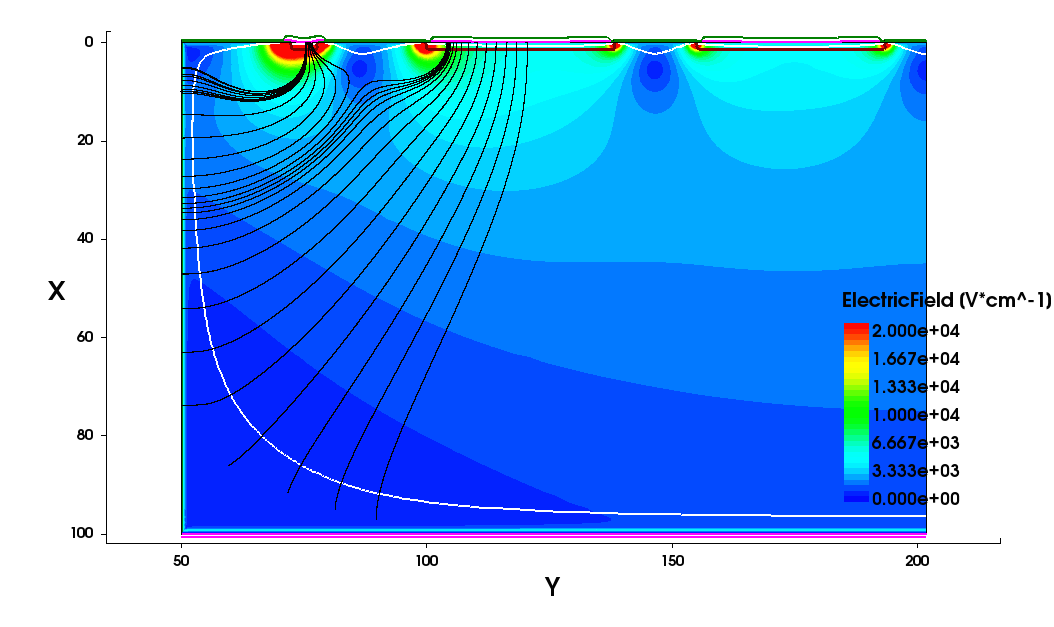}};
      \begin{scope}[x={(image.south east)},y={(image.north west)}]
      \end{scope}
    \end{tikzpicture}
    \caption{\SI{100}{\micron} thick}\label{fig:field_100um}
  \end{subfigure}
  \begin{subfigure}[b]{.31\linewidth}
    \begin{tikzpicture}
      \node[anchor=south west,inner sep=0] (image) at (0,0)
      {\includegraphics[width=.9\textwidth,clip=true,trim=6.5cm 2cm 12.5cm 0cm]{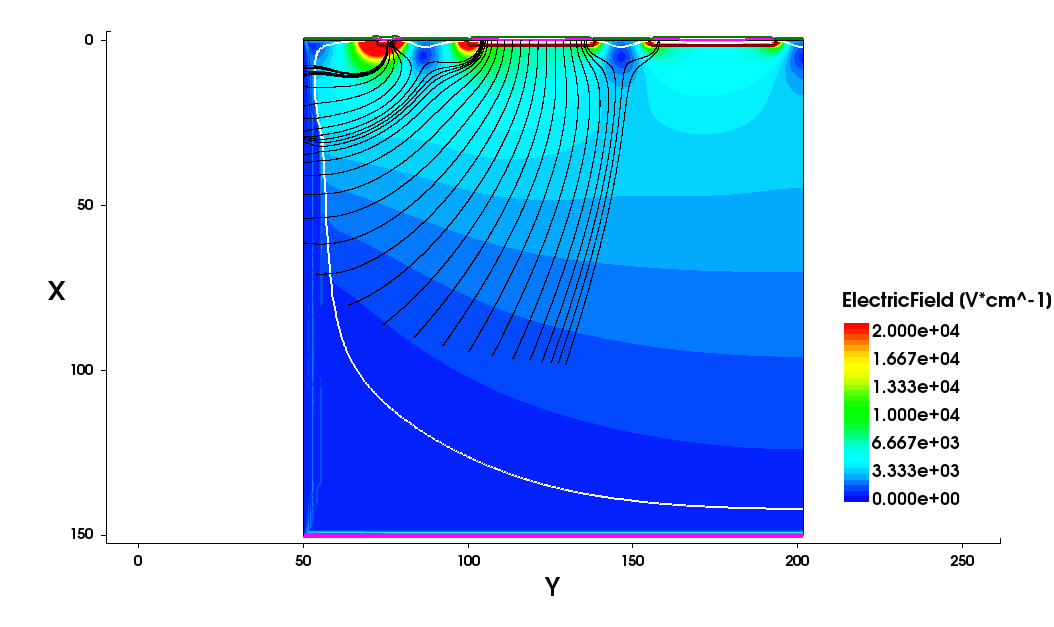}};
      \begin{scope}[x={(image.south east)},y={(image.north west)}]
      \end{scope}
    \end{tikzpicture}
    \caption{\SI{150}{\micron} thick}\label{fig:field_150um}
  \end{subfigure}
  \caption{Electric field distribution close to the sensor edge in (\subref{fig:field_50um}) \SI{50}{\micron} (at \SI{-15}{\volt}), (\subref{fig:field_100um}) \SI{100}{\micron} (at \SI{-20}{\volt}) and (\subref{fig:field_150um}) \SI{150}{\micron} (at \SI{-30}{\volt}) thick active-edge sensors with grounded guard ring and \SI{55}{\micron} edge distance. The solid white line indicates the border of the depleted volume.}\label{fig:efield_thickness}
\end{figure}

The influence of the sensor thickness on the charge collection efficiency at the edge is further clarified by investigating the electric field for the three different sensor thicknesses. The simulated field distributions are shown in \cref{fig:efield_thickness}. In the \SI{50}{\micron} thick device, most field lines reach from the edge directly to the grounded guard ring. Therefore, most of the charge created in the edge is collected by the guard ring and is lost for particle detection. In addition, the non-depleted volume near the sensor edge close to the backside adds to the charge loss.

In the edge region of the thicker device, only the field lines originating close to the sensor surface reach to the grounded guard ring. From a significant sensor volume deeper in the bulk the field is pointing to the pixel electrode, and thus charge deposited deeper in the sensor bulk is still collected by the first pixel. This enables the device to detect particles passing close to the physical sensor edge still fully efficiently, while it also explains the observed signal loss in the edge region of the thicker devices.

\section{Conclusions and outlook}
Thin planar silicon sensors with active edge and various guard ring layouts have been investigated using two-dimensional finite-element T-CAD simulations and with Timepix3 ASICs in test-beams. The detection efficiency as well as the charge collection efficiency have been found to depend strongly on the interplay between sensor thickness, guard ring position and guard ring connection scheme. The simulation results have been compared to experimental data, and an overall good agreement has been achieved. It has been demonstrated that \SI{50}{\micron} thick planar silicon sensors with active edge with floating guard ring or without guard ring can be operated fully efficiently up to the physical edge of the sensor. To prevent early breakdowns due to high field regions in devices without guard ring, a floating guard ring can be placed around the pixel matrix, without a significant impact on the detection efficiency. For application in the CLIC vertex detector, \SI{50}{\micron} thick sensors with narrow edge design and floating guard ring as well as the design without guard ring present a viable solution to achieve fully efficient particle detection up to the physical sensor edge with sufficient margin to the sensor breakdown. If necessary, further optimisation of the guard ring distance to the pixel implants could be performed. Prototype sensors with \SI{25}{\micron} pixel size have been produced in a similar active-edge process and are currently under test in combination with CLICpix2 ASICs~\cite{clicpix2-manual}.

\section*{Acknowledgements}
The authors thank Fernando Duarte Ramos (CERN) for his support  with the mechanical integration of the tested devices in the telescope system.
The help from the staff operating the CERN SPS and the North Area test facilities is gratefully acknowledged.
This project has received funding from the European Union's Horizon 2020 Research and Innovation programme under Grant Agreement no. 654168.
This work has been sponsored by the Wolfgang Gentner Programme of the German Federal Ministry of Education and Research.

\bibliography{bibliography}

\end{document}